\documentclass[]{aa}
\usepackage{graphicx,subfigure,natbib,url}
\begin{document}
\title{Spectral irradiance variations: Comparison between observations and the SATIRE model on 
solar rotation time scales}

\author{Yvonne C Unruh\inst{1}, Natalie A Krivova\inst{2}, Sami K Solanki\inst{2}, Jerald W Harder\inst{3}, Greg Kopp\inst{3}}
        \offprints{Y C Unruh}
	\institute{Astrophysics Group, Blackett Laboratory, Imperial College London, SW7 2AZ, United Kingdom
\\
   \and 
	Max-Planck-Institut f\"ur Sonnensystemforschung, D-37191 Katlenburg-Lindau, Germany \\
   \and
	Laboratory for Atmospheric and Space Physics, 1234 Innovation Drive, Boulder, 
	Colorado 80303-7814, USA 
              }
\date{Received \today; accepted }

\abstract
{}
{We test the reliability of the observed and calculated spectral irradiance variations between 200 
and 1600~nm over a time span of three solar rotations in 2004.}
{We compare our model calculations to spectral irradiance observations taken
with SORCE/SIM, SoHO/VIRGO and UARS/SUSIM.  The calculations assume LTE and are based 
on the SATIRE (Spectral And Total Irradiance REconstruction) model. We analyse the 
variability as a function of wavelength and present
time series in a number of selected wavelength regions covering the UV to the NIR. We also 
show the facular and spot contributions to the total calculated variability.}
{In most wavelength regions, the variability agrees well between all sets of observations and 
the model calculations. The model does particularly well between 
400 and 1300~nm, but fails below 220~nm as well as for some of the strong NUV lines.
Our calculations clearly show the shift from faculae-dominated variability in the NUV to 
spot-dominated variability above approximately 400~nm. 
We also discuss some of the remaining problems, such as the low sensitivity of SUSIM and SORCE
for wavelengths between approximately 310 and 350~nm, where currently the model calculations
still provide the best estimates of solar variability. 
}
{}


\titlerunning{Solar spectral irradiance variations on solar rotation time scales}
\authorrunning{Unruh et al.}
\maketitle

\section{Introduction} 
%
\label{sec:intro}
The solar irradiance, or the solar flux received at the top of the Earth's
atmosphere, is known to vary over a large number of time scales, ranging 
from minutes to months and decades. The changes in the total solar output 
have been measured since 1978 \citep{willson88} and different composites 
of the measurements have been presented by \citet{frohlich98,willson2003}
and \citet{Dewitte2004}.
While the short-term (minutes to hour) variability is mainly
due to solar oscillations and granulation, the daily to decadal variability 
is attributed to the changes in the surface magnetic field combined
with the solar rotation that transports solar active regions into and
out of view. Indeed, \citet{krivova2003_cycle23} found that more than
90\% of the solar variability between 1996 and 2002 could be explained
by changes in the solar surface field. Similar conclusions were reached 
by \citet{wenzler2006} who reconstructed solar irradiance from Kitt Peak 
magnetograms covering the last 3 solar cycles. 

Solar variability is a strong function of wavelength: while solar 
output is small in the UV, the relative variability is more than one 
order of magnitude larger in the UV than in the visible.
Until very recently, the spectral dependence of the solar variability
had mainly been determined in the UV, in particular by the 
measurements taken by the SUSIM and SOLSTICE instruments onboard UARS 
\citep[see, e.g.,][]{floyd-et-al-2003a}. Information  
in the visible was restricted to the three colour channels of
the SPM instrument of SOHO/VIRGO \citep{frohlich95}, though degradation 
hampered the use of these data beyond timescales of the order of a few 
months\footnote{SPM data are available from 
\url{ftp://ftp.pmodwrc.ch/pub/data/irradiance/virgo/SSI/spm_level2_d_170496_06.dat}; see also \url{ftp://ftp.pmodwrc.ch/pub/Claus/SORCE_Sep2006/SSI_Poster.pdf}.}. 

The variability at most other wavelengths had to be inferred
using a variety of approaches, such as e.g., pioneered by 
\citet{lean89} who produced the first estimate of
solar-cycle variability over a large wavelength range. An
alternative approach was followed by \citet{unruh99lumi} who used 
facular and spot model atmospheres to calculate the flux changes due 
to magnetic features. \citet{fligge2000irrad} and \citet{krivova2003_cycle23}
used solar surface images and magnetograms to calculate the 
variability on time scales
ranging from days to years. Here we built on this approach and 
present comparisons between modelled and measured spectral irradiances
during three months in 2004.

Thanks to missions such as SORCE and SCIAMACHY the observational outlook has 
now become much better and we have, for the first time, variability observations
that span from the UV to the near IR \citep{harder2005calib,rottman2005sorce,skupin-et-al-2005}. In the following we consider SORCE data only.
First comparisons between SORCE measurements and models have been 
presented by, e.g., \citet{fontenla2004sorce} and \citet{lean2005sorce}.

All data presented here have been recorded between 21 April and 1 August 2004.
During this time the Sun was in a relatively quiet phase, especially in May
when only a very small spot group appeared on the solar disk. A new and larger
active region emerged over the next month, resulting in a depression of just
over 1 permille in total solar irradiance (TSI) in July.

In the next section we briefly describe our irradiance modelling 
approach. We then discuss the data analysis for the different 
instruments (Sec.~\ref{sec:data}). In Sec.~\ref{sec:comps}, we 
compare the relative irradiance changes derived from the models 
with a number of different data sets spanning a wavelength range 
from 200 to 1600~nm. In particular, we compare our model to data from 
SORCE/SIM, UARS/SUSIM, and SoHO/VIRGO. We conclude this section by 
presenting observed and modelled timeseries in a number of selected 
wavelength bands. A discussion of the results and conclusions
are presented in Sec.~\ref{sec:conclusion}.

\section{Irradiance reconstructions}
Here we restrict ourselves to a brief description of our approach to model 
solar irradiance (see \nocite{fligge2000irrad,krivova2003_cycle23} Fligge et 
al.~2000, Krivova et al.~2003 for a more detailed discussion). 
Essentially, we calculate the solar irradiance (or flux) by integrating
over the (pixellated) solar surface, accounting for the presence of 
dark (sunspots) and bright (faculae and network) surface magnetic 
features. The location of sunspots is obtained from 
MDI continuum images, attributing penumbra and umbra to those pixels with
contrasts of less than 0.9 and 0.6, respectively. Faculae and the 
network are identified 
by their excess magnetic flux on MDI magnetograms. As faculae are 
very small-scale features and typically do not fill an entire MDI 
pixel, we adopt a filling-factor approach, scaling the facular filling
factor (linearly) with the magnetic field strength measured from the 
magnetograms. The identification of the magnetic features is described
more extensively in \citet{fligge2000irrad} and \citet{krivova2003_cycle23}. 

The model has a single free parameter, $B_{\rm sat}$, which takes into
account the saturation of brightness in regions with higher concentration of
magnetic elements \citep[e.g.,][]{solanki84fluxtubes,solanki92,ortiz2002}.
$B_{\rm sat}$ denotes the field strength below which the facular contrast is
proportional to the magnetogram signal, while it is independent (saturated)
above that. From a fit to the VIRGO TSI time series \citet{krivova2003_cycle23} 
obtained a value of 280~G for $B_{\rm sat}$, which is used here unchanged. 

Once each pixel on the solar surface has been identified as either 
(part) facula, quiet Sun, umbra or penumbra, we can attribute a corresponding emergent 
intensity to it and then proceed to carry out the disk integration. 
Note that the emergent intensities have to be known as a function of 
limb angle for each of the components present on the solar surface. 
The wavelength resolution and available range for the final spectral irradiance 
is determined by the wavelength resolution and range of the limb-dependent
emergent intensities. 

We calculate the intensities from the SATIRE set of model atmospheres 
\citep{unruh99lumi}, using Kurucz' ATLAS9 program \citep{kurucz93}. The model 
atmosphere for the faculae and network was derived using 
FAL P \citep{fontenla99} as a starting 
point, while the quiet-sun is Kurucz' standard solar atmosphere and 
the sunspot umbra and penumbra models are stellar models 
at 4500~K and 5150~K also taken from \citet{kurucz93}. 
As the model atmospheres and intensities are derived under the assumption of 
LTE, we expect our irradiances to become unreliable below approximately 
300~nm \citep[see, e.g.,][]{unruh99lumi,krivova2006uv}. 

For the comparisons presented in this paper, the availability of MDI 
images\footnote{All MDI images were obtained from the MDI homepage
at \url{http://soi.stanford.edu/}.} and groups of five consecutive 
magnetograms (which were averaged to 
reduce the noise) was reasonably good and we were mostly 
able to calculate irradiances on a 12-hourly interval. There are, however, 
some data gaps, most noticeably at the end of June with only three sets of images
with poorer quality between 2004 June 22 and June 30. 
%
\section{Solar irradiance observations: May to July 2004}
\label{sec:data}
The main instruments used for the comparisons are the spectral and 
total irradiance monitors from SORCE, SIM and TIM, respectively. These 
data are complemented by contemporaneous observations from UARS/SUSIM 
and VIRGO/SPM. In this section, we briefly describe the instruments 
used and discuss the data analysis.

SORCE (Solar Radiation and Climate Experiment) was launched in January 2003 and 
started science operations in March of that year. It is the first satellite to 
provide reliable daily measurements of the spectral irradiance variability for 
wavelengths longer than 400~nm. It  carries four instruments, all of which 
have been described in \citet{rottman2005sorce} and sources referenced therein. 

\subsection{SORCE/SIM}
\label{sec:sorce}
SIM primarily measures spectral irradiance between 300 and 2400~nm with an 
additional channel to cover the 200 to 300~nm wavelength region. We consider  
data taken with three of its five detectors, namely the UV detector ($200 - 308$~nm), 
the VIS1 detector (VIS1: $310 - 1000$~nm), and the IR detector ($994 - 1655$~nm). 
These will be discussed briefly in the following sections. We discard the 
data from the second visible-light detector as it suffers from both temperature and 
radiation-induced variability that cannot be fully removed. 
We were unable to use the longer-wavelength data
recorded by SIM/ESR as they were too noisy over the 
time span considered here. For more information on the design and 
calibration of SIM we 
refer to \citet{harder2005design,harder2005calib}. 

The results presented here are based on Version~10 of the SIM data 
reduction. The availability of SORCE data in the time considered here is 
reasonably good with only some data gaps and correction problems during two weeks 
around the end of June and beginning of July. The SOHO/MDI suffered from poorer
imaging data during these two weeks as well, making comparisons at these times 
more difficult. 

All three detectors provide measurements of the solar irradiance as a function of 
wavelength on approximately 12-h intervals. As discussed in \citet{harder2005calib}, 
SIM typically has 6 samples per resolution element, yielding an (un-aliased)
oversampling by about a factor of 2. In order to compare the data to the 
model calculations, we characterise them in two ways. Firstly, we consider the 
variability, treating each wavelength bin as an independent time series. As a 
measure of the variability, we adopt the standard deviation, calculated according to
\begin{equation}
\sigma(\lambda_i) = \sqrt{\frac{\sum_{j=1}^{n}
		    \left(f_j(\lambda_i) - \bar{f}(\lambda_i)\right)^2}{(n-1)}}, 
	\label{eq:var}
\end{equation}
where $\bar{f}(\lambda_i)$ is the mean flux at wavelength $\lambda_i$
and $f_j(\lambda_i)$ is the flux at time $j$ and wavelength $\lambda_i$.
So as to better illustrate the relative changes in each wavelength bin, 
we only plot the {\em normalised} standard deviation, i.e., 
$\sigma(\lambda_i)/\bar{f}(\lambda_i)$. This measure has the advantage 
of simplicity and universality, but has the disadvantage that it makes 
disentangling facular and spot variability difficult. Secondly, we look at 
time series in a number of selected wavelength bins. These include the 
VIRGO/SPM filter bands, which allow us to compare our model, 
SORCE/SIM and VIRGO/SPM with each other, and 
a number of bands that stand out in the variability plots. 

Before calculating the final RMS spectra, the mean spectra and the time 
series, we removed obvious outliers in the SIM data. This was done for 
each wavelength bin individually, by removing data points that deviated 
from the mean by more than $k \sigma$. The cutoff factor, $k$, was 
varied with wavelength to account for the different aspects of the 
faculae-dominated variability in the UV and the spot-dominated 
variability at longer wavelengths. We thus applied a symmetric 
cutoff in the UV, generally clipping data points more than 3.5$\sigma$ 
from the mean. In the visible and infrared, typically data points more than 
2$\sigma$ above and 4.5$\sigma$ below the mean were clipped. The clipped 
data were replaced by median values of the two previous and subsequent exposures. 

\begin{figure}
        \resizebox{\hsize}{!}{\includegraphics{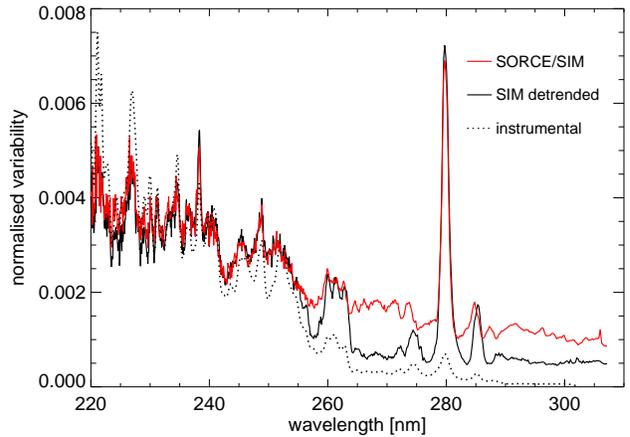}}
        \caption[]{
        The normalised standard deviation between 220 and 310~nm as derived
        from SIM/UV between 2004 April 21 and August 1. The red and black
	lines show the normalised standard deviation of the smoothed data
        after the removal of outliers. The difference between the 
	two variability spectra is due to the removal of a linear slope 
	that reduces the variability for wavelengths above about 260~nm
	as indicated by the black line (see section~\protect{\ref{sec:sorce}}
	for a description). The dotted line traces 
	the instrumental noise. Note that the instrumental noise 
	exceeds the binomially smoothed signal for wavelengths below approximately 230~nm. 
        }
\label{fig:uv_var}
\end{figure}

The variability plots are shown in Figs~\ref{fig:uv_var} and \ref{fig:vis1}
and will be discussed in the following sections. The plots also show the 
derived instrumental noise level.  Note that SIM is generally not photon-noise 
limited, but analog-to-digital converter (ADC) limited with about 2 bits of 
noise on a 15-bit converter range. It requires a dynamic range of aproximately 
100 to measure the signal, so for weak signals the noise level becomes comparable to the 
solar variability signature. Apart from these random noise contributions, additional 
residual systematic trends caused by the imperfect prism degradation and 
temperature corrections can still be present in the time series. As the analog-to-digital
noise is essentially random, the application of a (non-phase shifting) filter
is appropriate. Here we use binomial smoothing \citep{marchand1983} in the 
time domain with two passes of the lowest-order (1,2,1) filter, meaning that 
we will be insensitive to variability on time scales shorter than about 1.5~days. 

%
\subsubsection{SIM/UV}
\label{sec:instru_sim_uv}
Solar variability is very much higher in the UV than in the visible and 
infrared, and robust measurements with variations of the order of several 
percent are expected. \citet{harder2005calib} have shown that 
the response of the UV instrument becomes more unreliable towards the blue 
end of the wavelength range. Fig.~\ref{fig:uv_var} shows the normalised 
standard deviation for the SIM/UV data. The red line indicates the 
normalised standard deviation of the original Version~10 data set, once 
outliers have been removed. The black dotted line indicates the instrumental noise. 

Over the time span considered here, the data from the UV detector show
a slow, almost linear, decrease that introduces substantial variability and
can be picked up in the 264 to 277~nm and 290 to 300~nm regions in particular.
This decrease could be either instrumental or instrinsically solar in which 
case it would imply a slow decrease of the normalised solar UV irradiance 
at the 5000 ppm level over a 3-month time span. A solar origin is supported by
a comparison to the SORCE/SOLSTICE instrument \citep{snow2005}
over the same time span. While the SOLSTICE trend deviates in the first couple
of weeks, it generally agrees with the SIM measurements for the remainder 
of the time. The red line in Fig.~\ref{fig:uv_var} shows the normalised
standard deviation when the linear trend is removed. In this case, 
the variability around 270~nm decreases by a factor of 2 in better 
agreement with what is seen in the models (see Sec.~\ref{sec:mod_comps}).

The measured standard deviation agrees well between the smoothed and unsmoothed
data for wavelengths larger than approximately 260~nm where it also exceeds the 
instrumental noise by more than a factor of two. For wavelengths below 
240~nm the instrumental noise becomes comparable to the data variability. 
This indicates that the measured variability is largely instrumental on the 
roughly 12-hourly timescales considered here, and can thus be significantly 
reduced by binomial smoothing as indicated by the red line on 
Fig.~\ref{fig:uv_var}. Indeed, for wavelengths below 235~nm, the variability 
of the smoothed data falls below the instrumental noise.
%
\subsubsection{SIM/VIS1}
\label{sec:SIM_vis}
\begin{figure}
        \resizebox{\hsize}{!}{\includegraphics{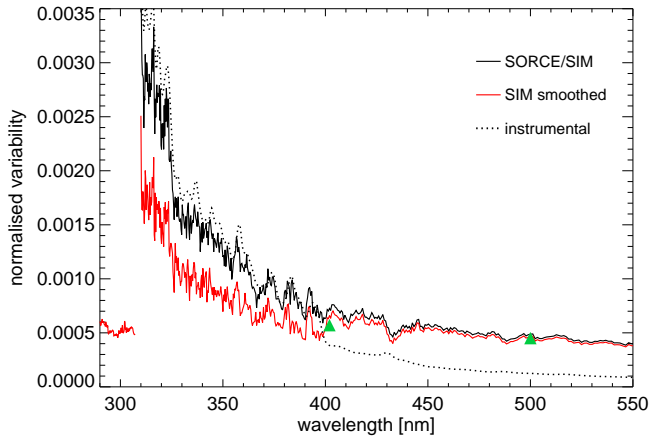}}
        \caption[]{Variability between 2004 April 21 and Aug 1 as 
        recorded with the SIM/VIS1 instrument.
        The black line shows the normalised deviation of the Version~10 
	data after the removal of outliers. The red line is for the 
	variability of binomially smoothed data, while the dotted 
	line indicates the instrumental noise. Also shown is the 
	variability measured in the VIRGO blue and green filters 
	(green triangles). The short data stretch below 308~nm is 
	as measured with SIM/UV.
	}
\label{fig:vis1}
\end{figure}

Fig.~\ref{fig:vis1} shows the variability measured with VIS1 for 
wavelengths between 310 and 550~nm. The red and 
black lines show the smoothed and unsmoothed data, respectively. 
Also shown is the instrumental noise level (dotted line). 
The figure illustrates that the instrumental variability increases 
dramatically for wavelengths shortward of approximately 400~nm 
\citep[see also figure 2 in][]{woods2007sns}. 
While there is no marked difference between the smoothed 
and unsmoothed data above 400~nm, indicating that we measure 
a predominantly solar signal, the variability of the smoothed 
data is lower (by a factor of about 1.5) at shorter 
wavelengths and falls below the instrumental noise level. 
It is thus not straightforward to estimate the solar variability
from the data available here, or indeed even estimate unambiguously 
the range up to which the smoothed data represent solar rather than 
instrumental signal. 

Overall, an increase in the standard deviation is expected for lower 
wavelengths, though the variability seen between 310 and 350~nm is 
clearly too high. We can take some guidance from the standard 
deviation of 500~ppm recorded with the SIM/UV detector 
at 300~nm. We would, however, caution against interpolating the 
variability between 300 and 390~nm, despite the similar variability 
levels observed at both wavelengths. Not only does the region 
contain a number of intermediate-strength lines, it also coincides 
with the expected switch-over between the facular and spot-dominated 
regime on rotational time scales. Depending on the exact balance 
between facular brightening and sunspot darkening, both effects can 
almost cancel each other out. This would explain, e.g., why the 
variability is lower at 385 and 395~nm compared to 400~nm. 

Not shown on Fig~\ref{fig:vis1} is the VIS1 variability above 550~nm, 
as it is mainly featureless: it shows a slow decrease between 
550~nm to 800~nm where it ranges from 350~ppm down to about 270~ppm. 
For longer wavelengths ($>820$~nm) it shows an upturn. This variability 
increase was already 
noted by \citet{harder2005calib} who attributed it to the incomplete 
removal of temperature-induced variability in the instrument. 
The variability recorded by SIM/VIS1 is shown over the 
full wavelength range and discussed further in Sec.~\ref{sec:mod_comps} 
where it is compared to the model results.
%
\subsubsection{SIM/IR}
SIM/IR, the infrared detector records the solar spectrum between 850~nm and 
1.66~$\mu$m. The data in the IR suffer from occasional sudden data jumps in time. 
The data become particularly noisy at the detector edges. We thus only use 
data for wavelengths between 980 and 1600~nm in the following. In 
this wavelength range, the data are very uniform with a normalised 
standard deviation between 230 and 300~ppm. 

%
\subsection{SORCE/TIM}
With TIM, the Total Irradiance Monitor, the SORCE satellite also carries a solar 
radiometer to measure total solar irradiance. The TIM instrument has been described 
in detail by \citet{kopp2005design} and first results have been presented in 
\citet{kopp2005results}. The instrumental noise level is less than 2~ppm and the 
instrumental stability is corrected to $<$10 ppm/yr, so the TIM data require
no long-term gradient removal or high-frequency temporal filtering for
the analyses here using Version~5 data. 
The TSI measured by SORCE/TIM is about 5~W~m$^{-2}$ lower than the TSI 
measured by other radiometers in space, such as, ACRIM-III and 
VIRGO \citep{froehlich97}. The irradiance changes of TIM, however, 
agree extremely well with those of the other radiometers, not 
only over the three months considered here, but also over the 
whole life time of the SORCE mission. Here we use TIM as 
representative for the TSI. As we consider relative changes in TSI 
only, we have normalised the modelled and SIM-integrated 
data (see Sec.~\ref{sec:tsi_comps}) to the SORCE/TIM values.
%
\subsection{UARS/SUSIM}
The Solar Ultraviolet Spectral Irradiance Monitor (SUSIM) is a dual
dispersion spectrometer instrument  that operated from 1991 to
2005 \citep{brueckner-et-al-93}.
It was one of the 2 UV experiments on board UARS (Upper Atmosphere Research
Satellite).
SUSIM has been monitoring solar irradiance in the range from 115 to 410~nm
with a spectral resolution between 0.15 and 5~nm.
We use the daily level~3BS V22 data with sampling of 1~nm
(\citealp{floyd-et-al-2003b}, Floyd, priv. comm. 2006) 
available at ftp://ftp.susim.nrl.navy.mil.
Calibration of the changing responsivity of SUSIM's working channel was
done through a combination of measurements of four on-board deuterium
calibration lamps and solar measurements by less frequently exposed
reference optical channels \citep{prinz-et-al-96,floyd-et-al-98}.
The long-term uncertainty of irradiance measurements (1$\sigma$) is about
2--3\% at $\lambda >170$~nm, $\approx 5$\% at $\lambda =140-170$~nm and
increases to around 10--20\% at shorter wavelengths
\citep{woods-et-al-96,floyd-et-al-98,floyd-et-al-2003b}.
%
\subsection{VIRGO}
The VIRGO/SPM instrument onboard SOHO measures solar variability in three 
wavelength bands, centred on 402 (blue), 500 (green) and 862~nm (red) 
with bandwidths (FWHM) of 5.4, 5.0 and 5.7~nm, respectively. 
The data presented here
are level 1.7 daily averages and have been obtained from the SOHO data
archive. SPM measurements suffer from 
strong and non-linear degradation, so that stretches longer than one 
month need to be corrected carefully before they can be used for 
comparison purposes. To correct for the degradation, we 
divided the VIRGO/SPM data by VIRGO TSI data and fitted a 
quadratic function to 9 data points that coincide with times of low
solar activity. 
This essentially pins the long-term behaviour in the colour channels to 
that of the TSI during quiet-Sun phases. 

Note that a newer SPM data release has recently become available 
where most of the long-term degradation has been 
removed (Fr\"ohlich 2007, priv.~comm). A comparison between our corrected 
data with the new data release shows that the variability amplitudes 
agree very well. A small amount of 
(possibly spurious) long-term variability, however, remains
even in the new data set. Rather than carrying out a similar procedure as outlined
above, we decided to use the old, but corrected data set. 

%
\section{Comparisons of the SATIRE model to SORCE/SIM and SOHO/VIRGO measurements}
\label{sec:comps}
\subsection{Total solar irradiance}
\label{sec:tsi_comps}

\begin{figure}
        \resizebox{\hsize}{!}{\includegraphics{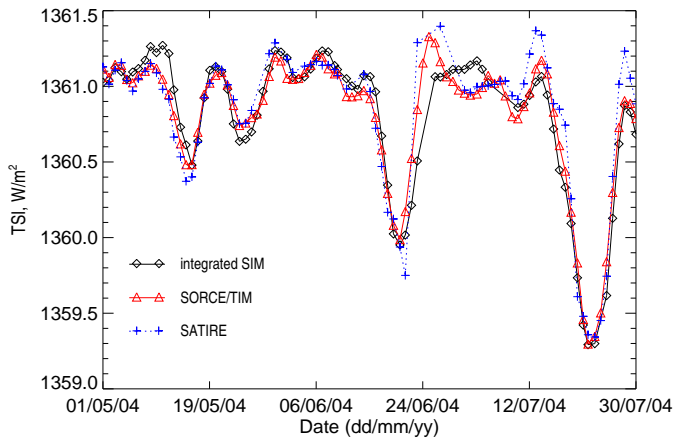}}
        \caption[]{Total solar irradiance (TSI) from May to July 2004. The black
                diamonds linked by the solid black lines show the integrated
                SORCE/SIM data after binomial smoothing; the red
                triangles show the SORCE/TIM total solar
                irradiance and the blue plus signs linked by the dotted
                lines indicate the integrated modelled irradiance.
                The model and SORCE/SIM data have been integrated between
                220 and 1660~nm and have been normalised so that their mean
                matches the absolute value of the mean SORCE/TIM TSI.
                }
        \label{fig:TSI}
\end{figure}

As a first test, we compare the modelled total solar irradiance to the
SORCE/TIM measurements as well as to the SORCE/SIM
`total' solar irradiance. This SIM pseudo-TSI was obtained by integrating the (smoothed)
SORCE/SIM measurements over the available SIM wavelength range. As this
does not cover the full solar spectrum, the resulting integrated irradiance
is, at 1230~Wm$^{-2}$, about 10\% lower than the TIM measurements. This
value is in reasonably good agreement with model expectations: we
find that the model irradiance between 220 and 1660~nm is 1207~Wm$^{-2}$.
The comparison between the time dependence of the modelled, SIM integrated
and TIM measured irradiance is shown in Fig.~\ref{fig:TSI}.
Both, the SIM wavelength-integrated data and the modelled TSI were
renormalised to match the absolute value of the TIM TSI. For the 
SIM-integrated TSI, renormalisation should yield an upper limit for the variability 
amplitude, as the missing part of the spectrum is mainly in the IR where 
variability levels are expected to be lower. We therefore also tried an 
approach whereby we added a constant offset. The true behaviour is 
then expected to lie between these extremes. We found both lightcurves
to be very similar with no significant changes for the correlation 
coefficients, and therefore only present the normalised lightcurves in 
the following.

\begin{table}
\caption[]{A list of correlation coefficients for the different TSI
determinations. The first two columns give the data sets used for the 
correlations, while the third column lists the Pearson linear 
correlation coefficient $r$ and its square. The fourth and fifth 
columns give the more robust Spearman rank correlation,
$\rho$, and its corresponding probability of a chance correlation.}
\begin{center}
\begin{tabular}{llccc}
\hline 
set 1           & set 2 & $r$ \ $[r^2]$   & $\rho$ & prob \\ 
\hline  \\  [-2ex]
integrated SIM  & TIM   & 0.97 [0.94] & 0.86  & $9\times 10^{-26}$ \\
model           & TIM   & 0.97 [0.94] & 0.89  & $9\times 10^{-28}$ \\
integrated SIM  & model & 0.92 [0.84] & 0.72  & $1\times 10^{-20}$ \\
\hline
\end{tabular}
\end{center}
\label{tab:TSI_corrs}
\end{table}

The agreement between the SORCE/TIM, integrated SORCE/SIM measurements and 
the modelled TSI is good overall, as borne out by the correlation 
coefficients and chance probabilities listed in 
Tab.~\ref{tab:TSI_corrs} \citep[see][\ \ for more detail]{press86}. 
The difference between the model and the data is of the same 
order as the difference between the two data sets. 
Closer inspection of Fig.~\ref{fig:TSI}, however, shows that some 
inconsistencies remain. The model appears least reliable between June 21 
and June 30. This is mainly due to a lack of MDI magnetograms and 
continuum images, and the poorer quality of those magnetograms 
and images that are available. Furthermore, the model seems to overestimate 
the facular brightening associated with the spot passage in July. 
The reason for such an excess brightening could either be that 
our facular contrast calculations are too high for large active regions, or
it could arise through errors in the feature identification, e.g., if some 
of the spot/pore magnetic flux were wrongly attributed to faculae. This 
may occur in particular when active regions are near the limb. Finally, 
uncertainties in the LOS magnetic field correction can lead to the magnetic 
field strength and hence the contrast of the faculae being overestimated.

The main difference between the integrated SIM data and the other 
two data sets occurs during the passage of the two small spot 
groups in May and in the period just after the sunspot passage in June. 
Compared to either TIM or SATIRE, the integrated SIM data show a 
larger flux increase before the passage of the first spot group, 
followed by a stronger flux decrease during the passage of the 
second spot group. It is not clear what might have led to this 
difference, as the data do not appear particularly noisy or 
discontinuous. At the end of June, SIM fails to pick up the facular 
brightening after the sunspot passage. In fact, the whole period between 
the two SIM data gaps (around June 24 and July 10) shows a different behaviour 
than expected from the model or the TIM data. 
In Sec.~\ref{sec:virgo_comps}, we show that not all wavelengths
are equally affected by this problem. Integration over a bluer 
wavelength stretch, e.g., one that excludes wavelengths above 800~nm for 
VIS1, produces a flatter response during that time. The lightcurve is 
otherwise very similar and the resulting increase in the correlation 
coefficient is very slight, so that it has not been plotted here. 
%

\subsection{Comparisons between SIM, VIRGO and the SATIRE model}
\label{sec:virgo_comps}
\begin{figure} 
        \resizebox{\hsize}{!}{\includegraphics{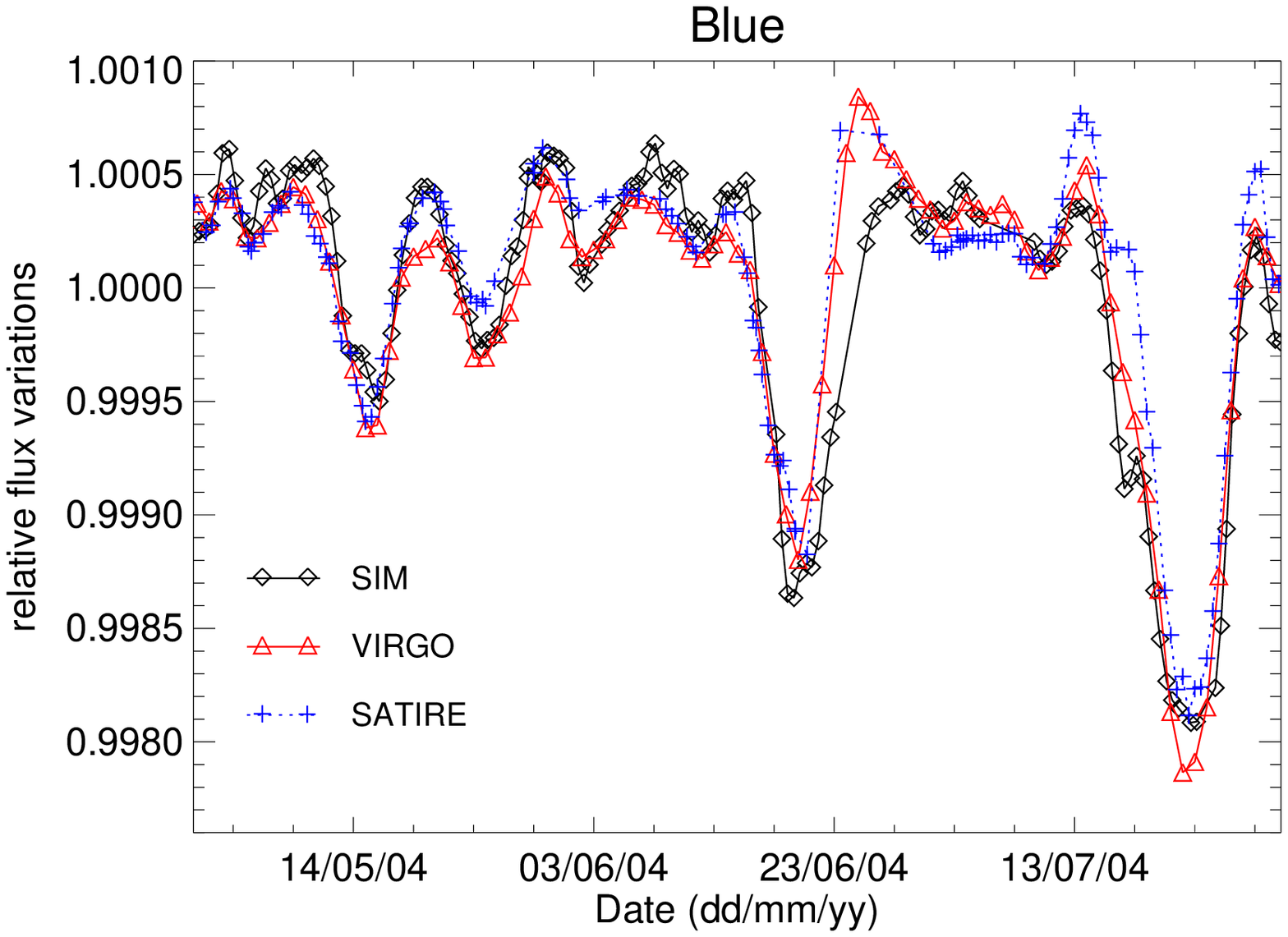}}
	\resizebox{\hsize}{!}{\includegraphics{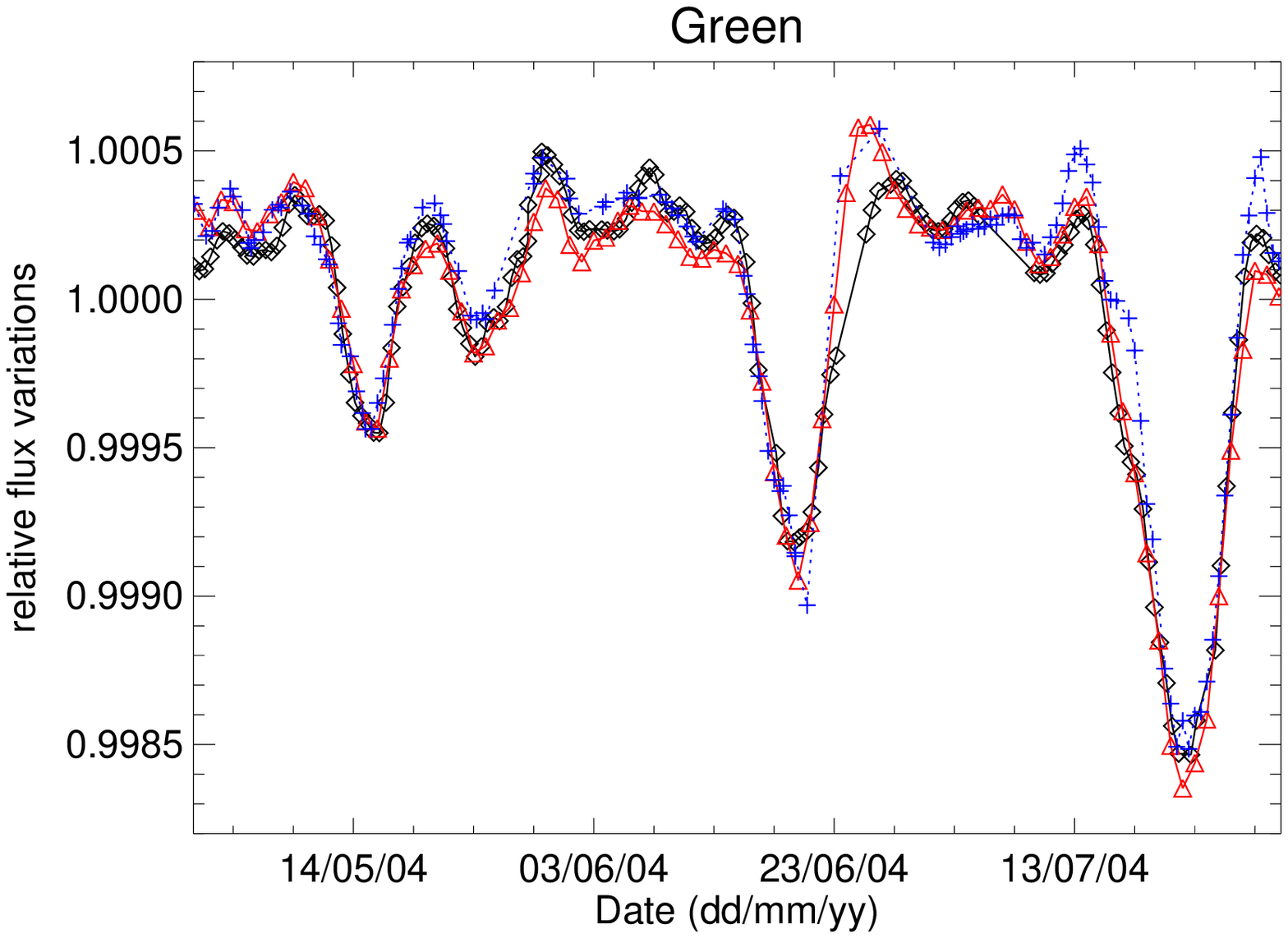}}
	\resizebox{\hsize}{!}{\includegraphics{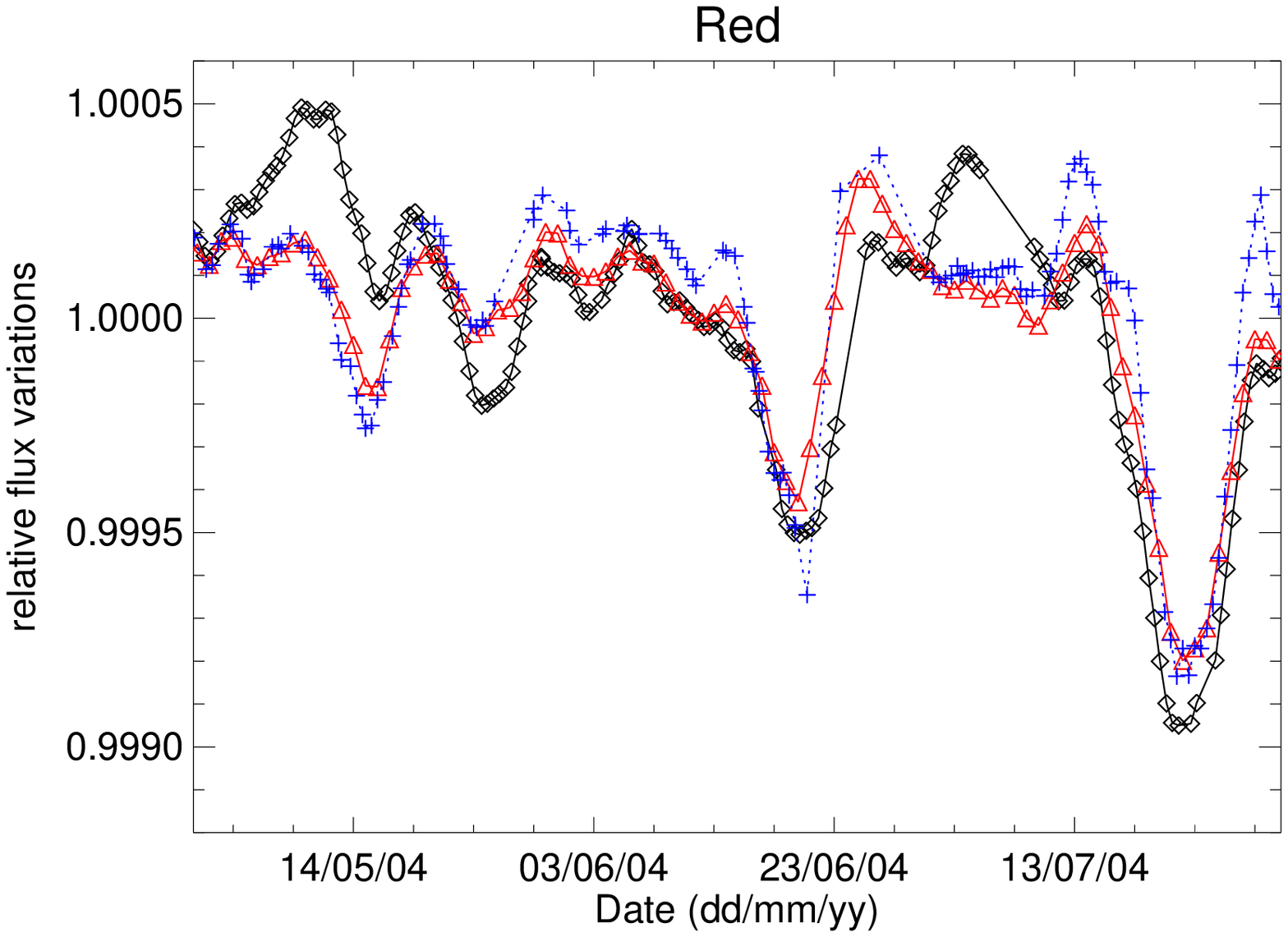}}
	\caption[]{Comparisons between detrended VIRGO/SPM data (red
		triangles and lines), and SORCE/SIM (black diamonds) as 
		well as model data (blue plus signs linked by dotted 
		lines) integrated according for the blue, green and 
		red VIRGO filters. In the bottom plot, the SORCE/SIM 
		data are for the red$_S$ filter (see text). Note the 
		different scales for the y-axis on the three plots.}
	\label{fig:VIR_comp}
\end{figure} 

Comparisons of VIRGO short-term spectral variations and our model have
been presented by \citet{fligge98virgo,fligge2000irrad} and
\citet{krivova2003_cycle23}.
Here we extend this work and compare model calculations to SORCE/SIM as
well as to the VIRGO/SPM irradiances. The VIRGO/SPM filters are narrow, 
to the extent that the FWHM of the green and red filters lie below the
corresponding SIM spectral resolution. The use of a detailed filter
profile is meaningless in such a situation and simply employing a single
SIM wavelength channel leads to overly noisy narrow-band fluxes. In order
to be able to include a larger number of wavelength bands and achieve
better signal-to-noise ratios, we used wider, rectangular filters,
ranging from 490 to 510~nm in the green, and from 830 to 900~nm in
the red. In the blue, we used the VIRGO/SPM bandwidth, as the number of
wavelength points covered by the blue filter is significantly larger
than in the green and red; furthermore, extending the blue filter is
difficult as there is not much clean continuum either
side of it. The filter widths and central wavelengths as applied to the
different data sets are listed in Tab.~\ref{tab:filters}.
\begin{table}
\caption[]{Central wavelengths and filter widths. For the VIRGO filters
largest and smallest wavelength indicate the extent of the FWHM. The
green and red filters for the model and SIM data are simple rectangular filters.}
\begin{tabular}{lcccc}
\hline
filter & \multicolumn{2}{c}{VIRGO} & \multicolumn{2}{c}{SIM/SATIRE} \\
       & $\lambda_c$ & range   & $\lambda_c$   & range \\
\hline
blue    & 402.6 & 400.0 -- 405.3 & 402.6 & 400.0 -- 405.3 \\
green   & 500.9 & 498.5 -- 503.3 & 500.0 & 490.0 -- 510.0 \\
red     & 863.3 & 860.5 -- 866.0 & 865.0 & 830.0 -- 900.0 \\
red$_S$ &       &                & 775.0 & 750.0 -- 800.0  \\
\hline
\end{tabular}
\label{tab:filters}
\end{table}

As discussed in Sec.~\ref{sec:SIM_vis}, the SIM/VIS1 detectors suffer from 
an imperfectly corrected temperature response above approximately 820~nm. 
This proved particularly troublesome in May 2004 and also during the 
data gap in early July. In order to exclude systematic effects arising through 
this, we also tested an alternative red channel (red$_S$, for short red) that
was obtained by integrating between 750 and 800~nm. This red$_S$ channel
suffers less from temperature-induced variability. Note, however, that the 
response in the two red channels is different. Our model calculations 
suggest that the variability in the shorter channel (red$_S$) should 
exceed that of the original VIRGO channel by approximately 10\%. 
Comparing the SIM data also suggests a larger response of the 
shifted red$_S$ channel with respect to the original VIRGO channel 
(by approximately 6\%), though this measurement is uncertain as, on 
account of the superimposed spurious variability, it has to be determined 
from a shorter data train.

Comparisons between the VIRGO/SPM data, SIM and our model are shown in 
Fig.~\ref{fig:VIR_comp}. We find that a number of days stand out 
in all filters. As already found for the TSI, the strongest discrepancy 
regarding SIM data occurs around June 25, just after the June sunspot 
passage. SIM apparently fails to pick up the facular brightening as 
the active region is near the limb; this is particularly salient in 
the blue and green filters. Note that the red filters mirror some 
of the problems seen in Sec.~\ref{sec:tsi_comps}, namely an 
enhancement in mid May just before the first spot group appears and 
a rise at the end of June before the data gap. These are even more 
pronounced in the original red filter (data not shown here) and we thus 
conclude that they are largely due to an incomplete removal of the 
instrumental temperature changes that affect the longer wavelength 
regions particularly strongly. 

\begin{figure}
        \resizebox{\hsize}{!}{\includegraphics{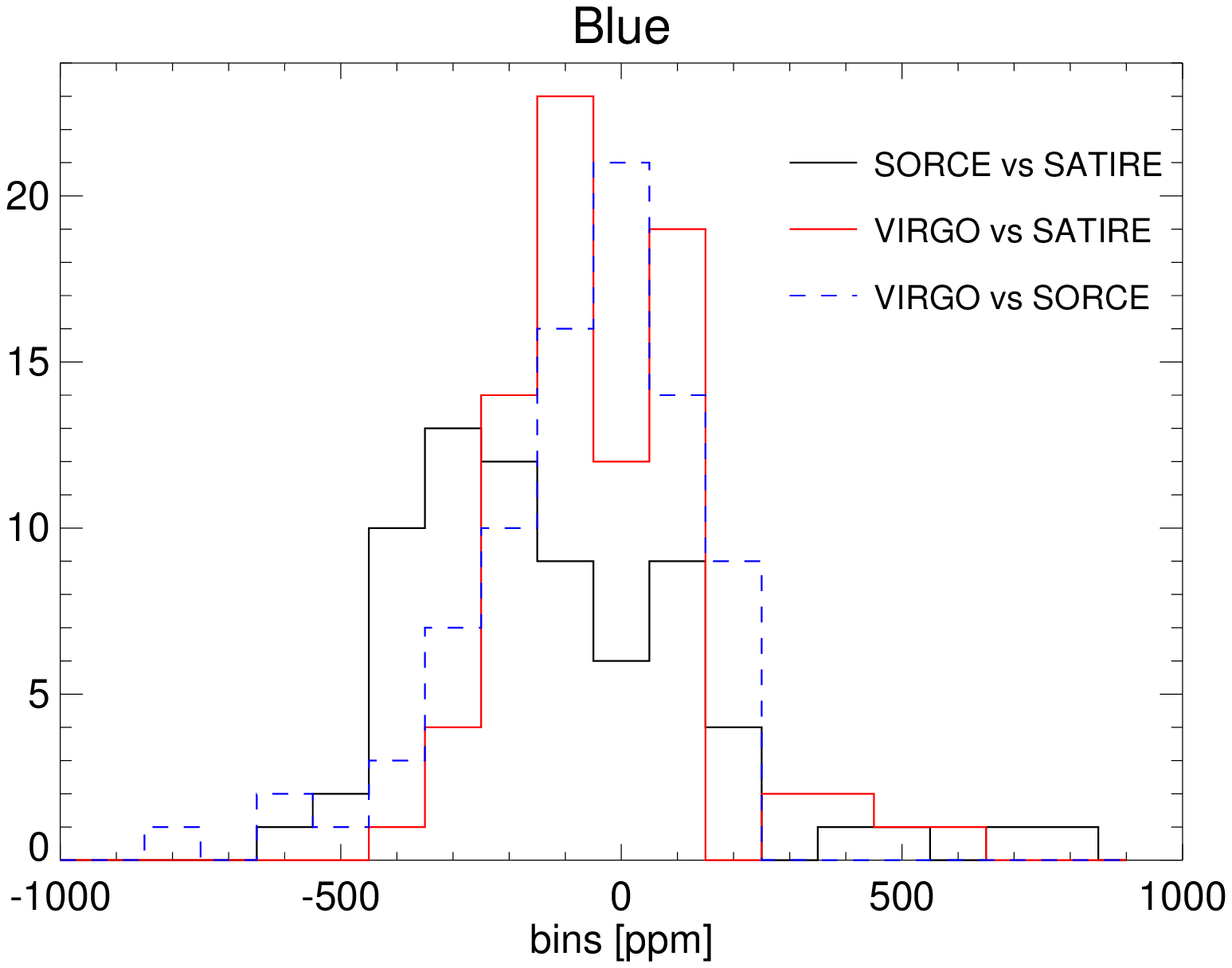}}
        \resizebox{\hsize}{!}{\includegraphics{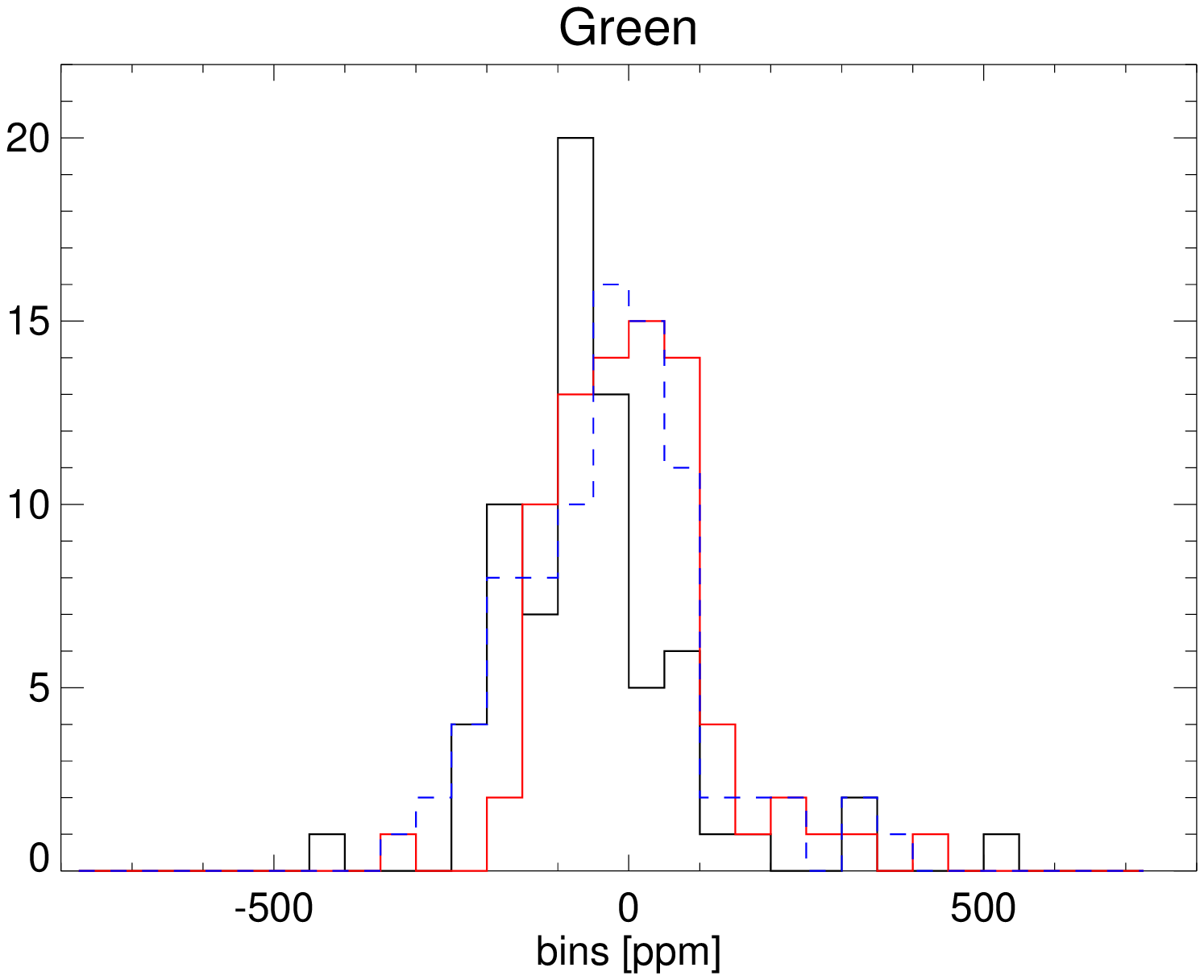}}
        \resizebox{\hsize}{!}{\includegraphics{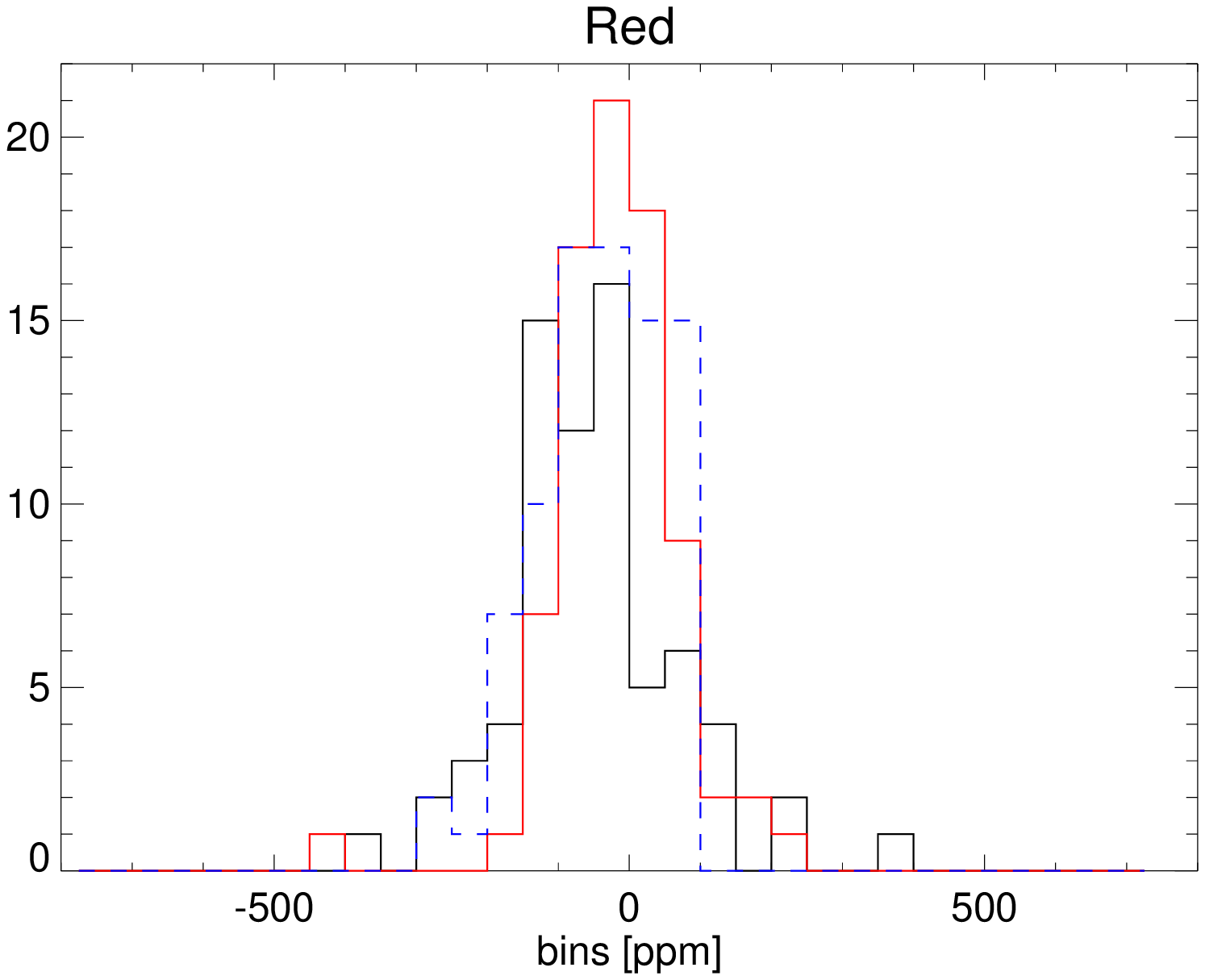}}
        \caption[]{Histograms illustrating the differences between
                the measurements and the model in the three VIRGO/SPM
                channels. The black and red histograms compare SIM, 
		respectively SPM, against SATIRE. The blue dashed lines 
		trace the histograms for SPM vs SIM. Note that bin widths 
		for the blue filter have been doubled; and that the shorter
                red$_S$ filter was used for the SIM comparisons.
                }
        \label{fig:histos}
\end{figure}

The largest difference between the model and both SIM and VIRGO data is
the much larger facular brightening before and after the
July sunspot passage. This is very noticeable in all three filters, as 
indeed also for the TSI (see Sec.~\ref{sec:tsi_comps}). By contrast, 
we find that the  sunspot darkening measured by VIRGO and SIM agrees very 
well with the models, in particular in the green and blue filters; in 
the red band there is a slight tendency for the darkening to be 
overestimated.  An exception to the good fit is the very small spot 
at the end of May where the model underestimates the sunspot darkening 
in all filters.

One way to judge the tightness of the correlation between
two data sets is to consider histograms of their fractional differences
as shown in Fig.~\ref{fig:histos}. Because of the response problems
of SIM/VIS1 for wavelength in excess of about 820~nm, we have used the
short red filter (red$_S$) to obtain the fluxes for the SIM-to-model 
comparisons. The original central wavelength was used for the 
model-to-VIRGO comparison and the two different filters are used when 
comparing VIRGO to SIM. While the number of data points considered 
here is relatively small, we find that most of the histograms resemble 
Gaussians, some of them with noticeable skew. The largest deviations 
are seen for the blue filter for the SIM vs SATIRE comparison, where 
the distribution is very broad and may be double-peaked. In most of
the cases, however, we can use the width of the standard deviation of
a Gaussian fit to the histograms to characterise the scatter between the
different data sets. These are listed in Tab.~\ref{tab:virgo_correls}. 

\begin{figure}
        \resizebox{\hsize}{!}{\includegraphics{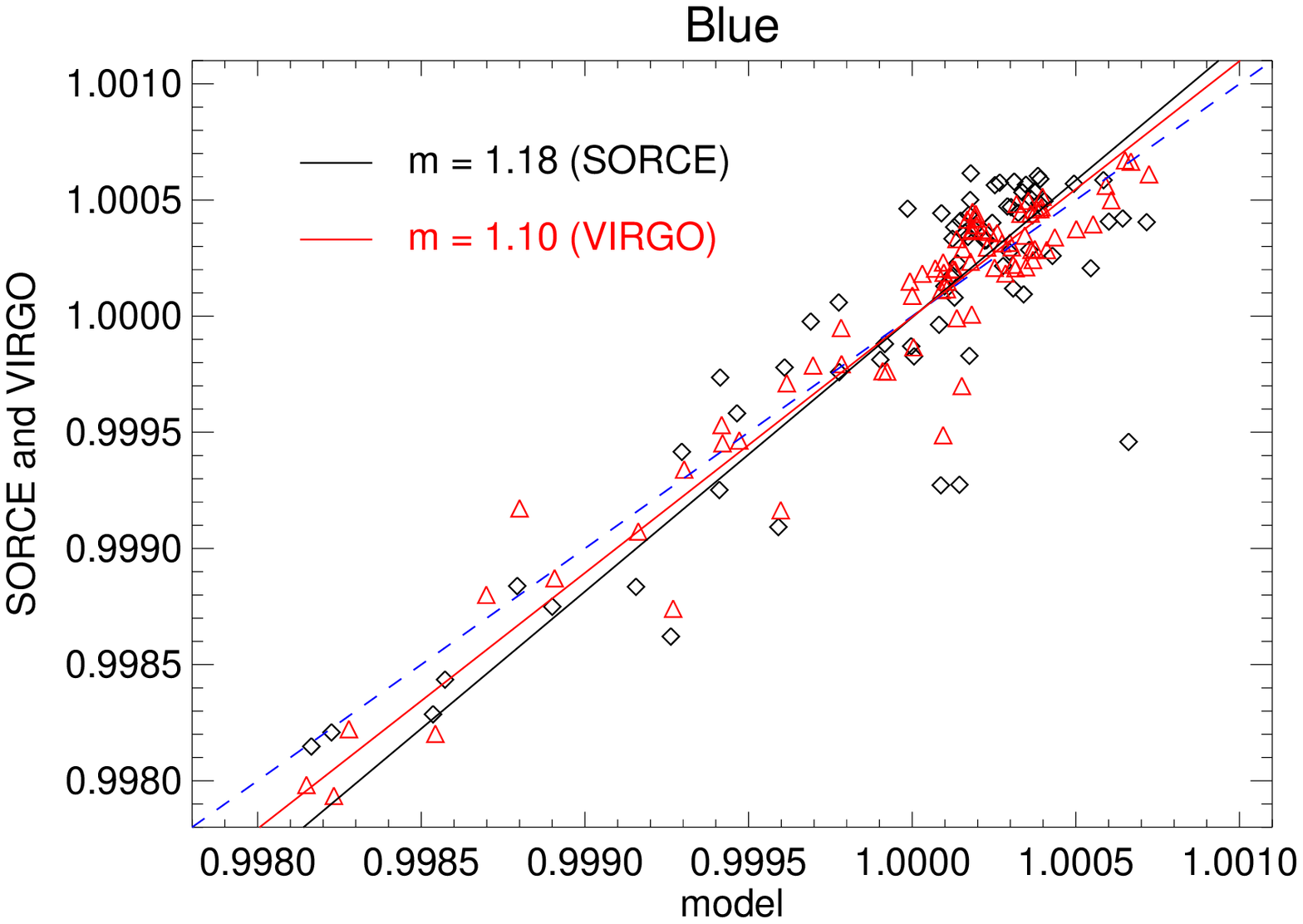}}
        \resizebox{\hsize}{!}{\includegraphics{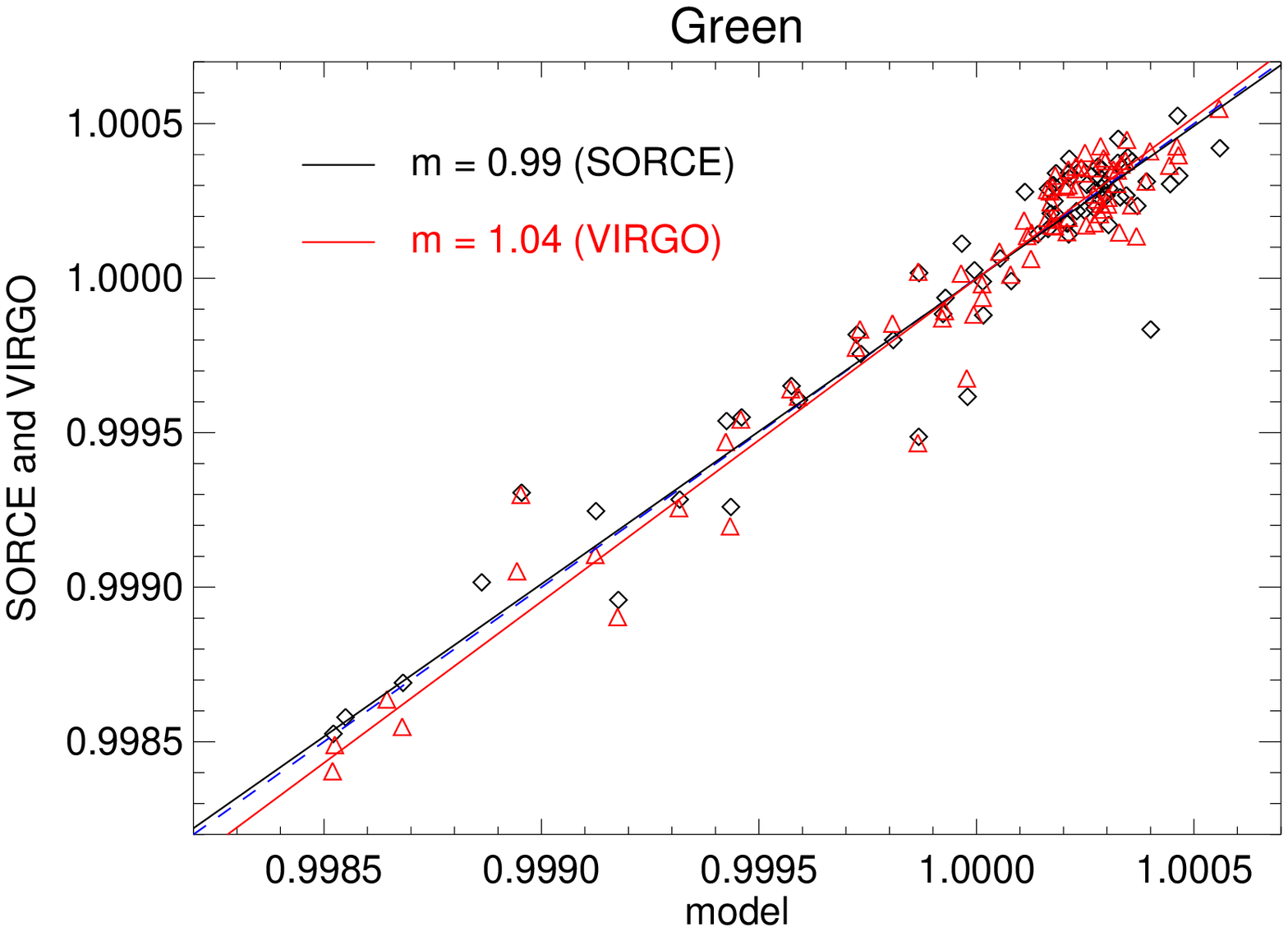}}
        \resizebox{\hsize}{!}{\includegraphics{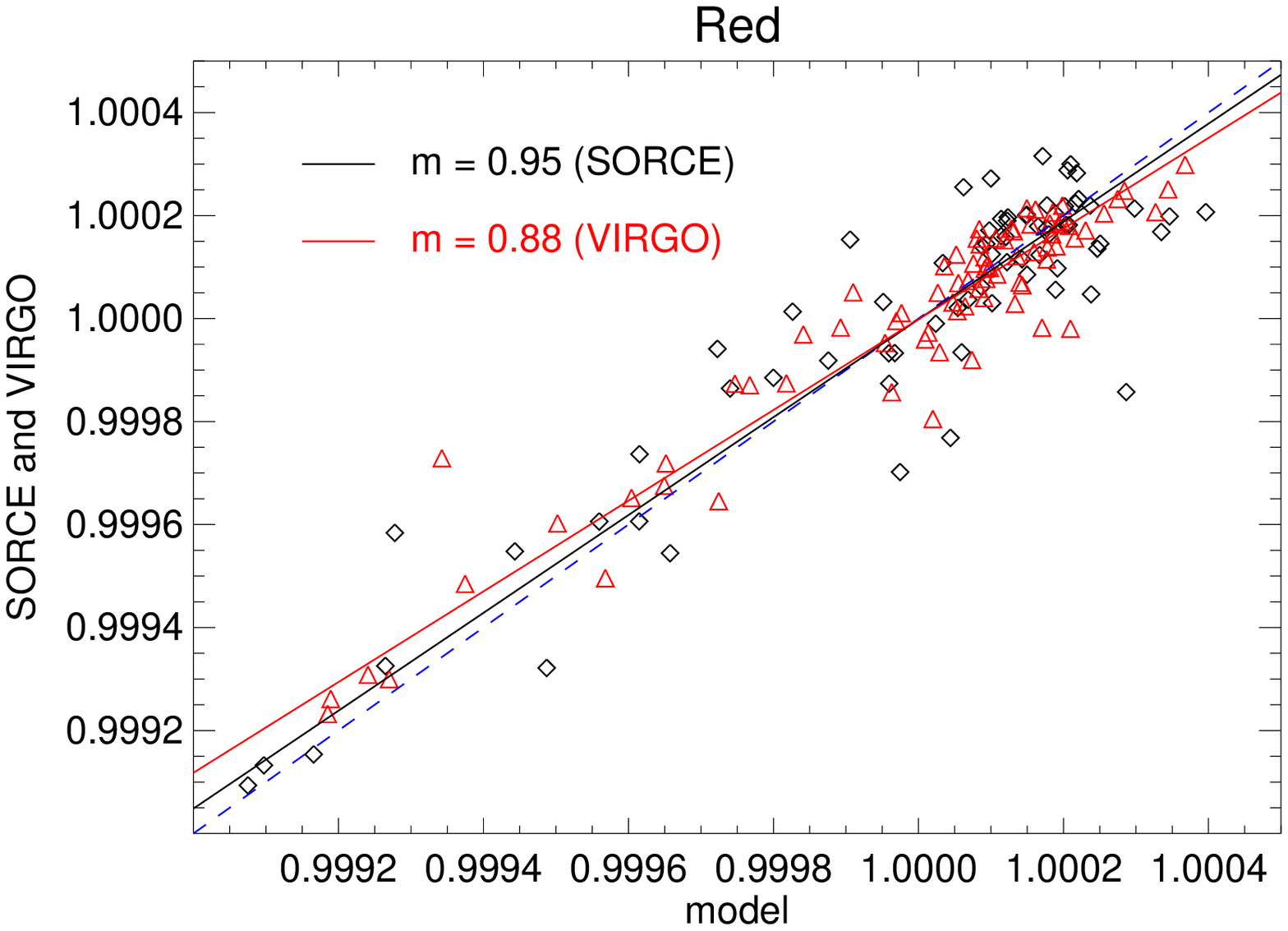}}
        \caption[]{Plots of VIRGO/SPM and SORCE/SIM vs SATIRE in the
                blue (top), green (middle) and red (bottom) VIRGO 
                channels. The black diamonds are for SIM vs 
                model, the red triangles for VIRGO vs model
                data. The blue thin dashed line indicates a unit
                gradient, while the solid black and red lines 
                show the gradients for the best linear fits for
                the SIM and VIRGO data, respectively.  }
        \label{fig:VIR_corr}
\end{figure}

In order to quantify the fits further, we list the correlation 
coefficients in Tab.~\ref{tab:virgo_correls}. To calculate the 
correlations, we binned the SORCE/SIM and model data onto the same 
grid as the daily VIRGO/SPM results. Fig.~\ref{fig:VIR_corr} 
shows the correlation plots of the VIRGO/SPM and the SIM filter 
observations with respect to the model calculations. The red and black 
lines show the best-fit lines, assuming that all data sets suffer from 
equal relative errors. These errors were estimated to be of the order 
of 100 ppm in the green and red filter, and of the order of 180 ppm 
in the blue filter (see also Tab.~\ref{tab:virgo_correls}). 
The dashed blue line indicates a slope of unity as
would be expected for a perfect match. 

When considering all filters together, the correlation coefficients and 
histogram widths suggest that the best agreement is found between the 
VIRGO data and SATIRE calculations, while comparisons fare least well 
for SIM versus SATIRE. The correlation coefficients and plots indicate 
that the agreement between the model and the observations is best for 
the green filter, but slightly less good for both the red and blue 
filters. In the following, we discuss the fits for the individual filters 
in more detail.

The correlation coefficients in the red filter show a relatively large 
variation, ranging from a tight correlation with coefficient 0.96 
($r^2$=0.90) between VIRGO data and model calculations, down to 
a coefficient of 0.77 ($r^2$=0.59) for SIM/red to model comparisons. 
This latter value can largely be explained by the temperature-induced 
sensitivity problems for SIM/VIS1 above about 850~nm. A comparison between 
the model and SIM data for the slightly shorter red$_S$ filter where this is 
less of an issue gives a significantly higher correlation coefficient
of 0.92 ($r^2$=0.84). Despite the filter shift and the associated change in 
responsivity to spot and facular passages, there is also an 
increase in the correlation coefficient (from 0.89 to 0.96) 
when comparing VIRGO red with SIM/red$_S$.  

In terms of the correlation coefficient, there is no significant difference 
between using the model calculations for the red or red$_S$ filter to compare
with a given observed time series. This is expected, as the basic 
features of the model differ only very slightly between the two 
wavelength bands. The amplitude of the variability, however, matches much 
better when the model calculations are carried out for the filter 
appropriate for the comparison data. 

The slopes derived for the red correlations show that SATIRE overestimates 
the amplitude of the variability by about 5\% (SIM) and 10\% 
(VIRGO). For the SIM analysis, we used the alternative `short' red filter 
as this is essentially unaffected by the temperature effects. 
Note that the very small gradient of 0.88 derived for the VIRGO vs model 
comparisons is in part due to a single outlying model point on June 21, 
shown as the red triangle at (0.99935, 0.99975) in the bottom plot of 
Fig.~\ref{fig:VIR_corr} (see also Fig.~\ref{fig:VIR_comp}). If this 
outlier is excluded, the gradient increases to 0.91 which is in better 
agreement with the results for SIM. A further reason for the low
gradient could be that the red filter band includes the redmost line of 
the Ca~{\sc ii} IR triplet that might not be modelled well by SATIRE.

The correlations between all the data sets are excellent in the green 
filter. In fact, they are better than those determined for the TSI 
comparisons, presumably because they are limited to the similar narrow 
spectral region and therefore sample a well-definied region of the 
solar atmosphere. We also find that the gradients of the best-fit lines 
are near unity, indicating that the amplitude of the variability agrees 
between all three datasets. In the blue filter, the agreement between 
the model calculations and VIRGO is comparable to that of the TSI 
comparisons, though the model fares less well with respect to the SIM 
data. As indicated by the correlation gradients, the model appears to 
underestimate the variability by between 10 and 20\%. We consider 
the 20\% derived from SIM data to be less reliable, mainly because the 
data follow a slight degradation-like long-term trend. This is again 
most obvious at the beginning of May. A correlation analysis with 
data after May 11 yields a slope of 1.11 with respect to the model, 
and agrees well with our findings for the blue VIRGO filter.

\begin{table}
\caption[]{Table listing the standard deviation to gaussian fits to the 
histograms ($\sigma_{\rm i}$, in ppm) the linear correlation coefficient
($r_{\rm i}$), its square, and the regression slope ($m_{\rm i}$)
in the three VIRGO/SPM filters. Columns 2 and 3 list the quantities for 
the comparison for VIRGO/SPM against SORCE/SIM and the SATIRE models, 
respectively, while column 4 compares the SORCE/SIM measurements
to the SATIRE model calculations. Standard deviations in curved 
brackets indicate that the histogram had significant skew and/or 
sidelobes and that a Gaussian was not a good fit. In the rows
with the regression slopes, the square brackets give the errors 
on the regression slopes, assuming equal errors for SIM, VIRGO/SPM 
and SATIRE. Note that there are no VIRGO data for the red$_S$ filter; 
the last three lines in columns 2 and 3 thus give correlation 
coefficients and slopes comparing the VIRGO red band and the 
SIM or model red$_S$ bands.}
\begin{center}
\begin{tabular}{lccc}
\hline
                		& VIR vs SOR    & VIR vs SAT & SOR vs SAT \\
N               		& 84            & 79            & 71 \\
\hline
$\sigma_{\rm blue}$             & 172       & 146           & (231) \\
$r_{\rm blue}, [r_{\rm blue}^2]$ & 0.95 [0.90] & 0.96 [0.92] & 0.89 [0.79] \\
$m_{\rm blue}$ 			& 1.07 [0.04] & 1.10 [0.05] & 1.20 [0.06] \\ [1ex]
$\sigma_{\rm green}$            & 109       & 95           & 92   \\
$r_{\rm green},[r_{\rm green}^2]$ & 0.98 [0.96] & 0.97 [0.95] & 0.96 [0.92] \\
$m_{\rm green}$ 		& 0.78 [0.05] & 1.04 [0.03] & 0.99 [0.04] \\ [1ex]
$\sigma_{\rm red}$              & (85)       & 71           & (177)   \\
$r_{\rm red},[r_{\rm red}^2]$ 	& 0.89 [0.78] & 0.95 [0.90] & 0.77 [0.59] \\
$m_{\rm red}$ 			& 1.46 [0.04] & 1.14 [0.06] & --  \\ [1ex]
$\sigma_{\rm red,s}$            & 93        & --            & 91   \\
$r_{\rm red,s},[r_{\rm red,s}^2]$ & 0.96 [0.92] & 0.96 [0.91] & 0.92 [0.84] \\
$m_{\rm red,s}$ 		& 1.28 [0.05] & -- &  1.04 [0.06]       \\ [1ex]
\hline
\end{tabular}
\end{center}
\label{tab:virgo_correls}
\end{table}

\subsection{Comparisons between SIM, SUSIM and the SATIRE model}
\label{sec:susim_comps}

\begin{figure}
        \resizebox{\hsize}{!}{\includegraphics{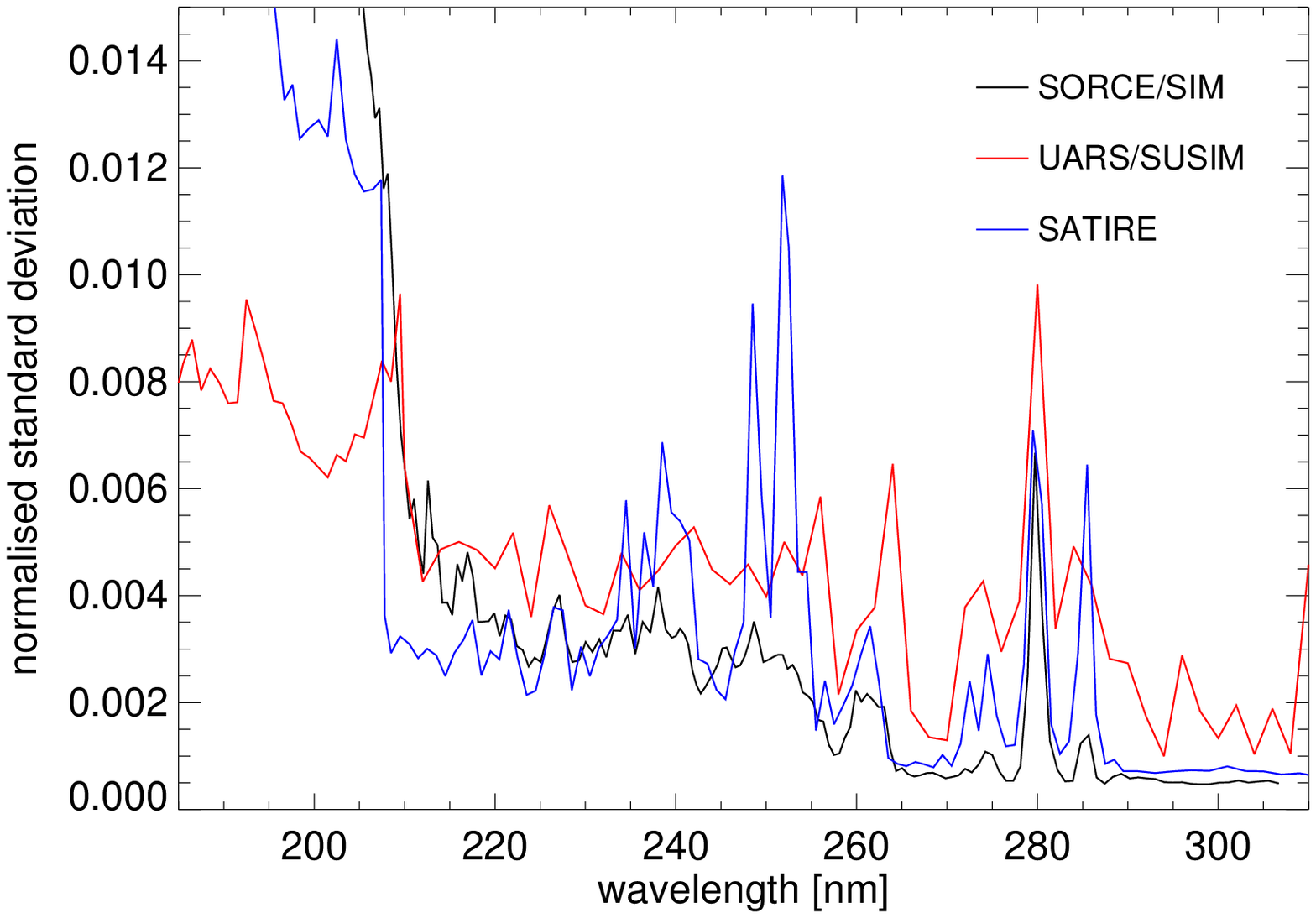}}
        \caption[]{Normalised standard deviation for SORCE/SIM (black line),
        UARS/SUSIM (red) and the SATIRE model (blue) calculations. To lessen
        the confusion of the plot, the SIM data have been wavelength binned
        by factors of 10 and 5 below and above 240~nm respectively; the 
        SUSIM data were binned by a factor of two for wavelengths 
	above 210~nm. 
        }
\label{fig:SOR_SUS}
\end{figure}

During 2004, the solar UV spectrum and its variability was also recorded with the
UARS/SUSIM instrument. In this section, we compare these measurements with
the SORCE/SIM measurements and the model calculations for wavelengths between
170 and 320~nm. Fig.~\ref{fig:SOR_SUS} shows a plot of the normalised standard
deviation for data and model calculations between May 1 and July 31 in 2004.
In order to reduce confusion in the plot, we have binned the SORCE/SIM and UARS/SUSIM
data in the wavelength domain before carrying out the variability analysis.
The binning factors are detailed in the caption of Fig.~\ref{fig:SOR_SUS}.
A number of striking features are apparent in the plot, and are discussed below.

The SORCE/SIM and SATIRE data show a large increase in variability 
at about 205~nm. This is most likely due to Al and Ca opacity edges 
between 200 and 210~nm. A significant decrease of the solar brightness 
temperature around this wavelength was already observed by 
\citet{widing70} based on data from rocket flights. It is very 
noticeable, however, that the variability recorded by UARS/SUSIM is 
much lower than both the SORCE/SIM and the SATIRE variability. 
While the UARS/SUSIM increase might appear weakened because of the 
higher (instrumental) variability seen above 200~nm and the lower 
velocity resolution, we expect the SUSIM data to best reflect the 
solar variability below 200~nm during the time period considered here.
The main reason for the (excessive) increase in variability of the 
SATIRE model is the breakdown of the LTE assumptions and the use of 
opacity distribution functions (ODFs) rather than detailed line-opacity 
calculations. In the case of SIM, the jump in variability is not 
surprising either, as its detector is not expected to perform well at 
these wavelengths (see Sec.~\ref{sec:instru_sim_uv}).
Better results should be provided by SORCE/SOLSTICE.

In the wavelength region between 210 and 290~nm, the 
agreement between the model calculations and the SIM measurements is
varied. While some features such as the Mg~{\sc ii} h\&k lines 
(280~nm),  and the regions from 220 to 232~nm, from 255 to 270 and
above 290~nm match well, other  
wavelength regions show large disagreements, see, e.g., the lines of 
Mg~{\sc i} at 285~nm, or the complex sets of lines around 240 and 250~nm.
As above, these disagreements are due mainly to the assumption of 
LTE and the use of ODFs \citep{kurucz92missing_uv}. Uncertainties in
the model atmospheres also contribute, though probably to a much smaller
extent, since the differences are largest at the wavelengths of 
strong lines showing strong NLTE effects. Thus the difference in
the behaviour of the Mg~{\sc i} and  Mg~{\sc ii} resonance lines
can be explained quite well if NLTE effects are taken into
account \citep{uitenbroek1995}.

Overall, the agreement between SIM and SUSIM is reasonably good between
about 210 and 290~nm. While the variability recorded with SUSIM is 
higher due to its lower sensitivity, the features recovered agree well
and the measured relative variabilities are not too different. 
Above 290~nm UARS/SUSIM shows a 
relatively poor response that swamps the solar variability on mid to
short-term time scales. The responsivity and noise characteristics
of SUSIM have been well documented and are discussed, e.g., in
\citet{woods-et-al-96}. 

%
\subsection{Comparison of SIM with the SATIRE model over the whole wavelength range}
\label{sec:mod_comps}
\begin{figure*}
        \resizebox{\hsize}{!}
        {\includegraphics{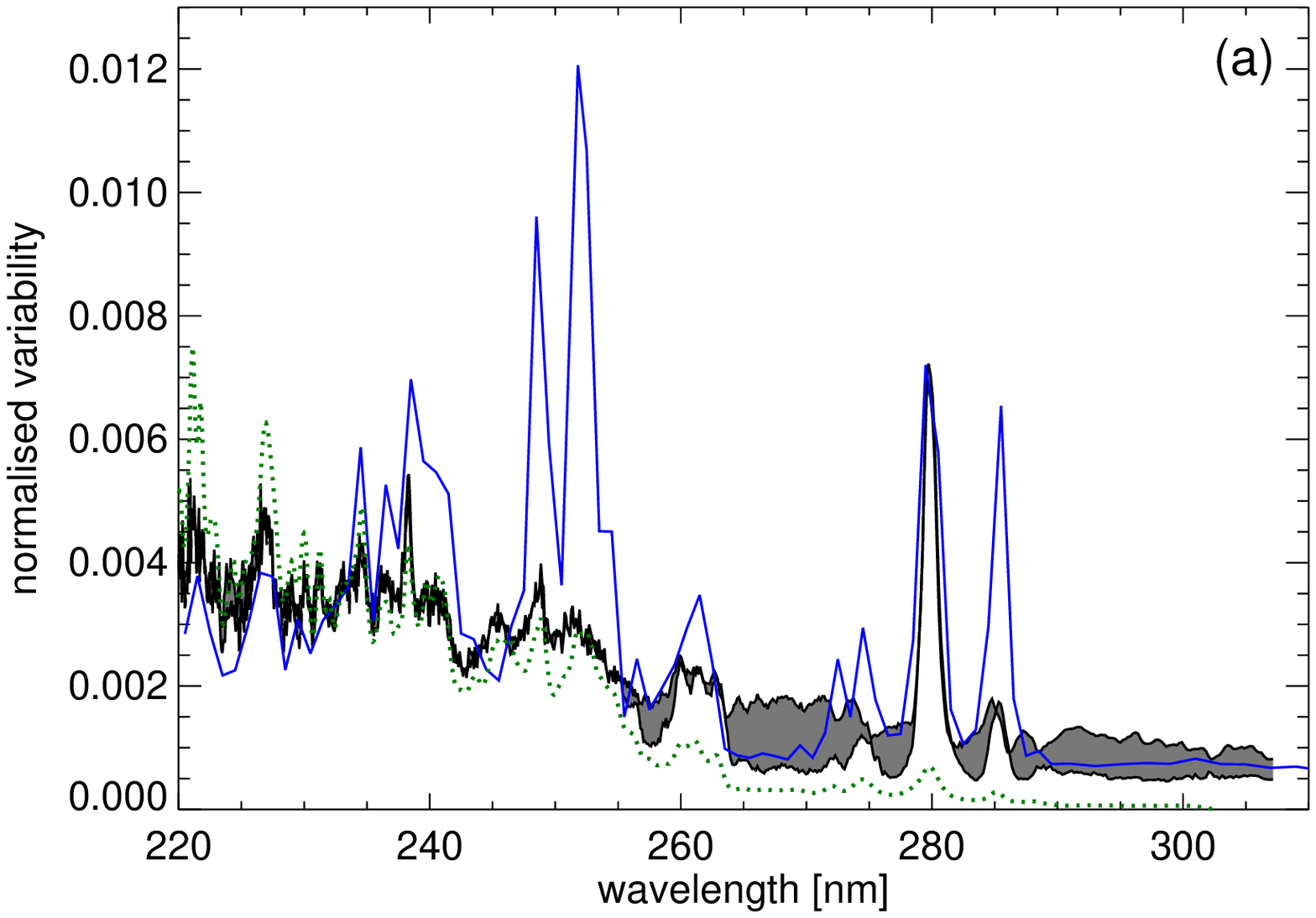} \includegraphics{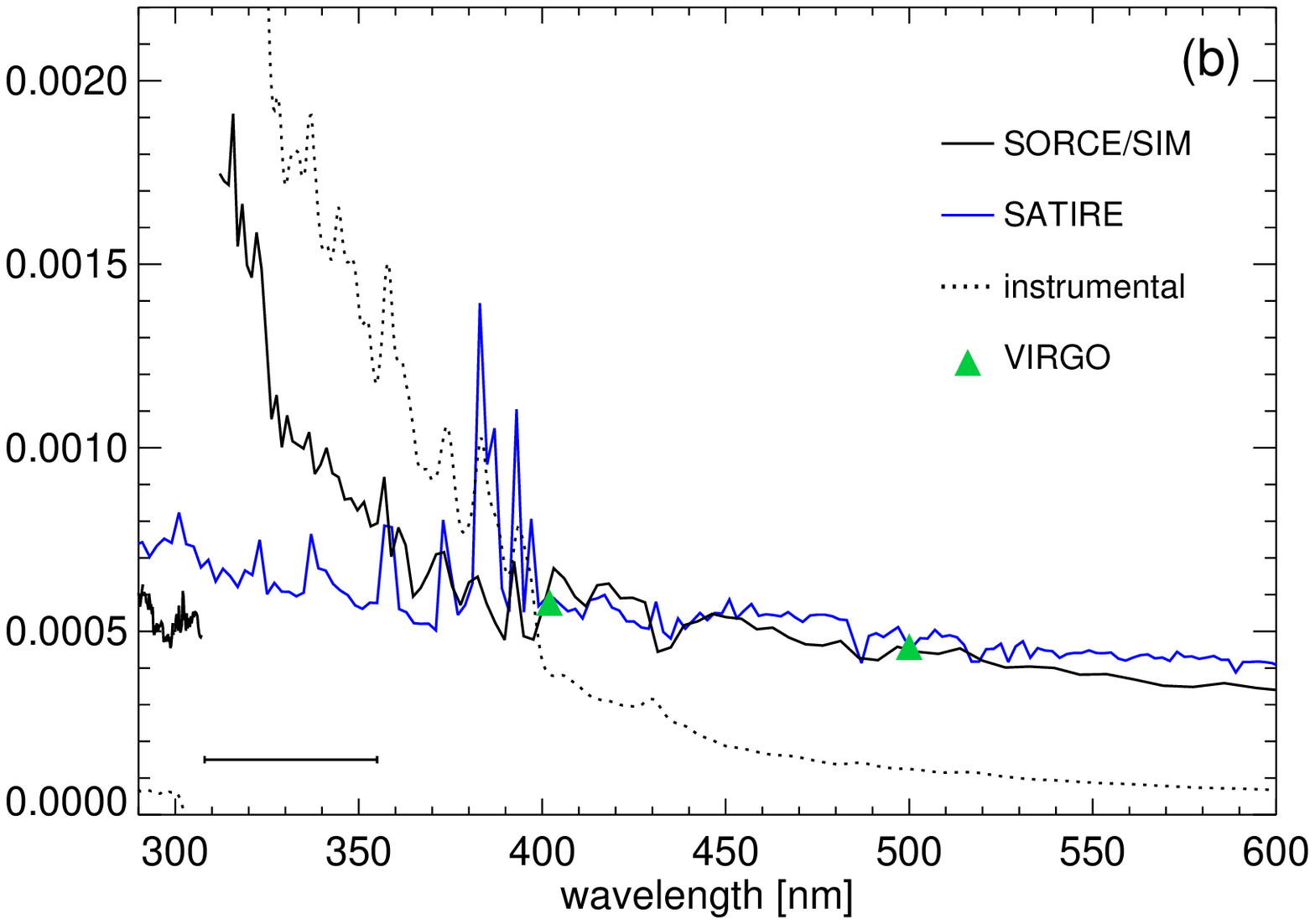} \includegraphics{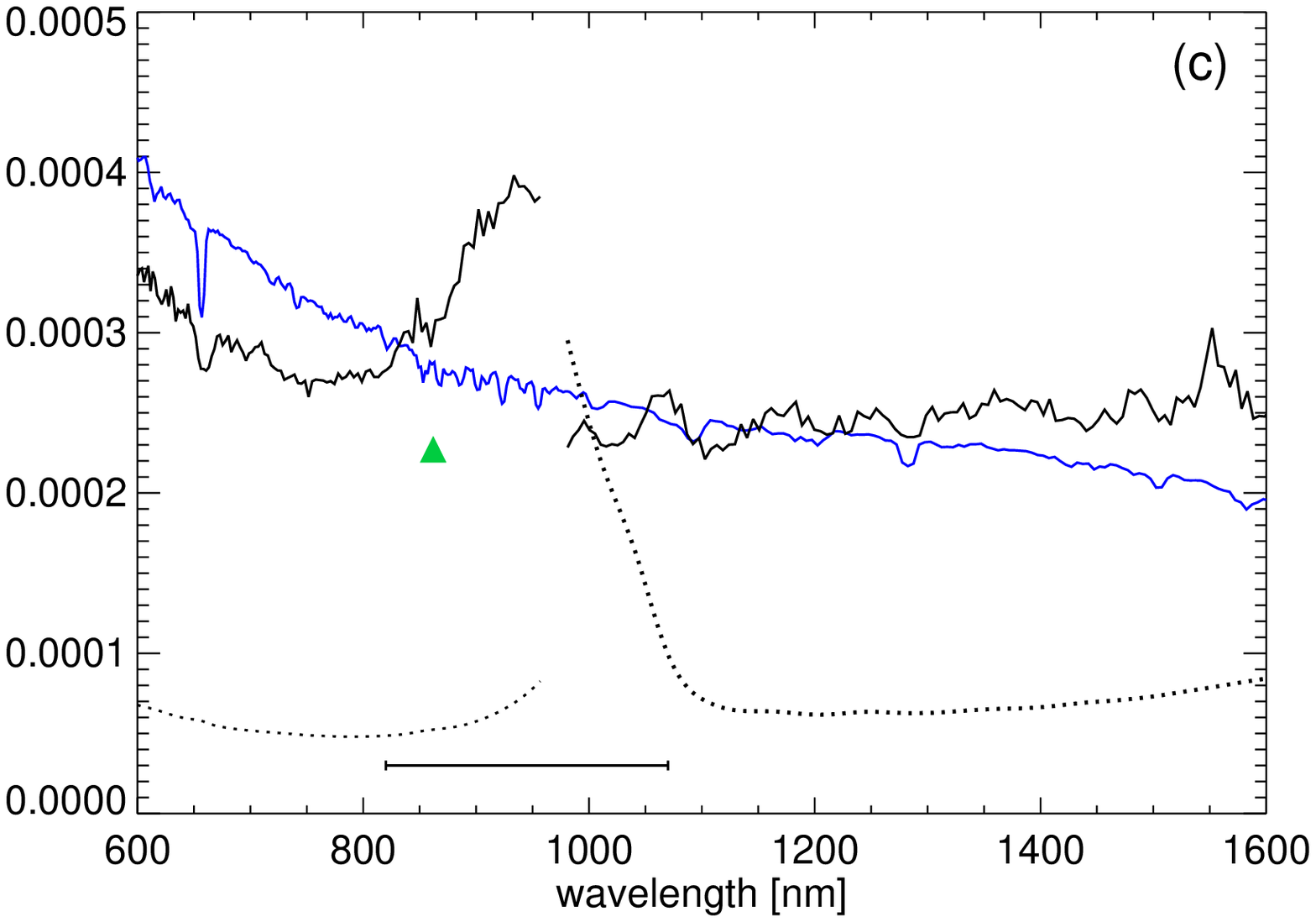}}
        \resizebox{\hsize}{!}
        {\includegraphics{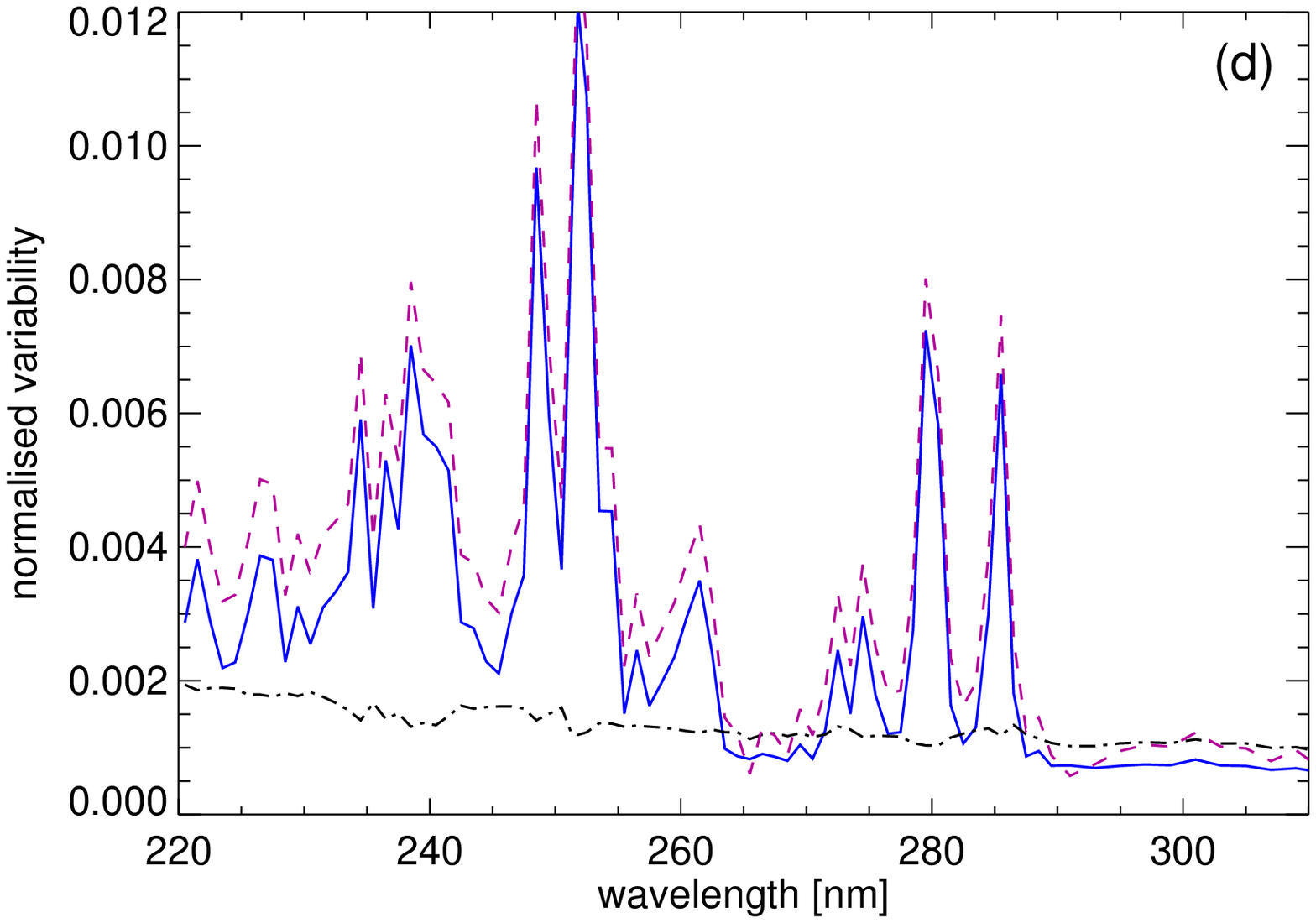} \includegraphics{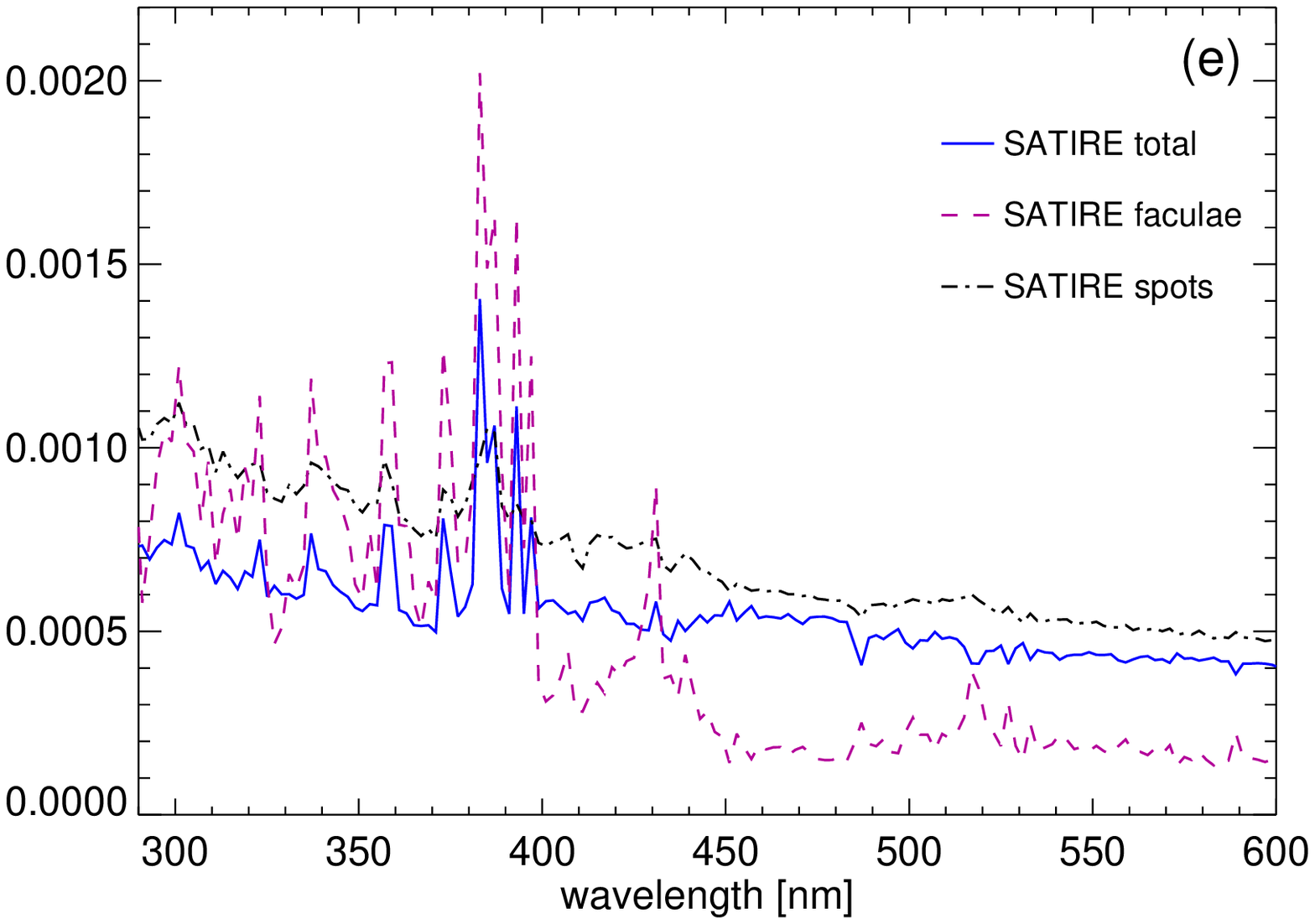} \includegraphics{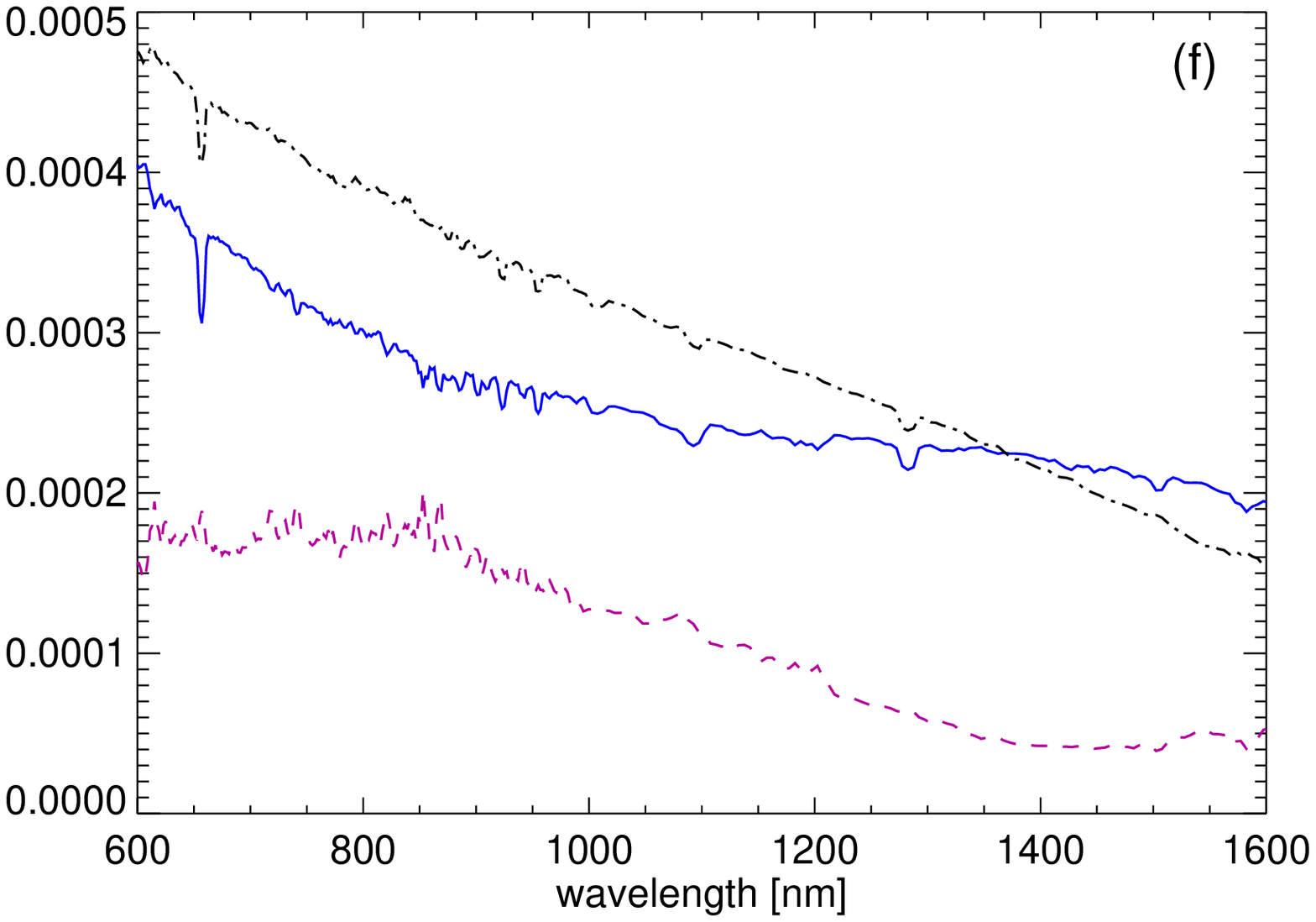}}
        \caption[]{Normalised standard deviation over the whole SORCE/SIM wavelength
        region. From left to right, the panels show the variability between 220
        and 320~nm, between 290 and 600~nm and between 600 and 1600~nm. The top
        figures show the modelled and the observed variability in blue and black,
        respectively; in the left-most plot, the grey-shaded area indicates the
        range of variability for SIM, depending on whether a linear trend is
        removed. Also shown is the instrumental noise limit (dotted lines) 
	and the variability measured in the three VIRGO filters (green 
	triangles). The horizontal solid lines offset from the plots indicate 
	the wavelength ranges where instrumental noise and artefacts dominate 
	the variability.
        The bottom plots show the modelled variability, distinguishing between
        total variability (blue solid line), the spot (dotted black) and facular
        (dashed purple) contributions.
        }
\label{fig:var_all}
\end{figure*}
The main advantage of SIM over the VIRGO/SPM channels is that a much 
more complete sampling of wavelengths is available. Fig.~\ref{fig:var_all} 
shows the RMS variability between May and August 2004 over the 
whole analysed SORCE/SIM wavelength range. The normalised standard deviation 
of the SORCE/SIM data is indicated by the solid black lines 
in panels a to c, though note that the grey-shaded area on panel a indicates
the range in variability that is obtained depending on whether a linear trend
is removed from the data or not. The horizontal bars in Figs.~\ref{fig:var_all} b and c
indicate the wavelength regions where the measured variability is dominated 
by instrumental noise (black dotted lines). The modelled variability is indicated 
by the blue lines. The bottom plots (d, e and f) show the contributions of the 
spots (dotted black lines) and faculae (dashed purple lines) to the overall 
variability. To calculate these, we replaced the facular (resp.~sunspot) 
contribution by a quiet-Sun contribution. The curves show very clearly 
that the wavelength dependence of spot variability is spectrally much 
smoother than the facular variability. This has to do with the fact that 
the darkening due to spots is dominated by the drop in continuum 
intensity. Changes in spectral lines produced by the lower temperature 
in spot umbrae and penumbrae play a secondary role. For faculae, the 
absolute temperature difference is less pronounced (especially in the 
lower atmosphere), and it is the different temperature gradient that 
produces changes in the continuum as well as in the lines. Especially at 
shorter wavelengths, individual and groups of lines provide the dominant 
contribution \citep{mitchell91,unruh2000issi}.

Below about 280~nm, the variability due to faculae 
generally exceeds that due to spots; then follows a region up to 400~nm 
where they are mostly of comparable magnitude. 
Above 400~nm, the modelled spot variability is always larger than the 
facular variability, and the combined variability follows the spot variability 
closely. Note that there is a further cross-over around 1.4~$\mu$m where the 
modelled variability of the spots and faculae drops below the total variability. 
This marks the transition where the facular model becomes dark averaged over 
the solar disk and thus no longer acts to counterbalance the spots. The fact that 
the model shows a smaller variability than SIM at these wavelengths cannot be due 
to this property of the model; dark faculae enhance the darkening due to spots, 
and thus increase the standard deviation (see also Sec.~\ref{sec:conclusion}). 
In other words, if the model faculae were bright at these wavelengths 
the discrepancy between the SATIRE model and SIM data would be larger. 

The shift in importance away from faculae to spots at wavelengths 
as low as about 300~nm is expected when considering variability on the 
order of a couple of month, i.e., on the rotation time scale when the 
influence of spots tends to dominate the TSI. On longer time scales 
such as that of the solar cycle, however, bright small-scale 
magnetic features dominate the TSI variations, and are thus also 
expected to dominate variability in both the UV and visible. 
\begin{figure*}
        \resizebox{\hsize}{!}{\includegraphics{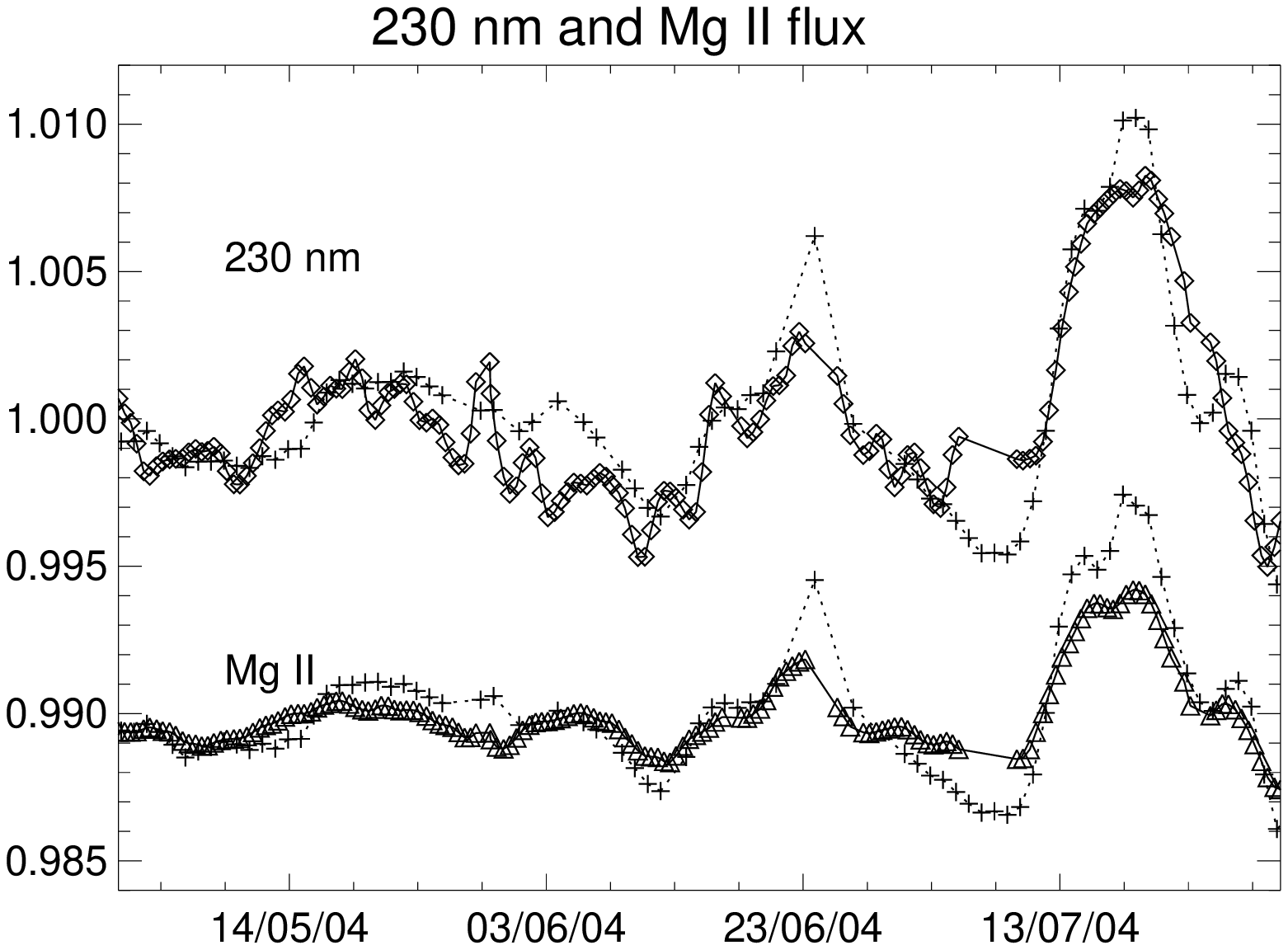}
                              \includegraphics{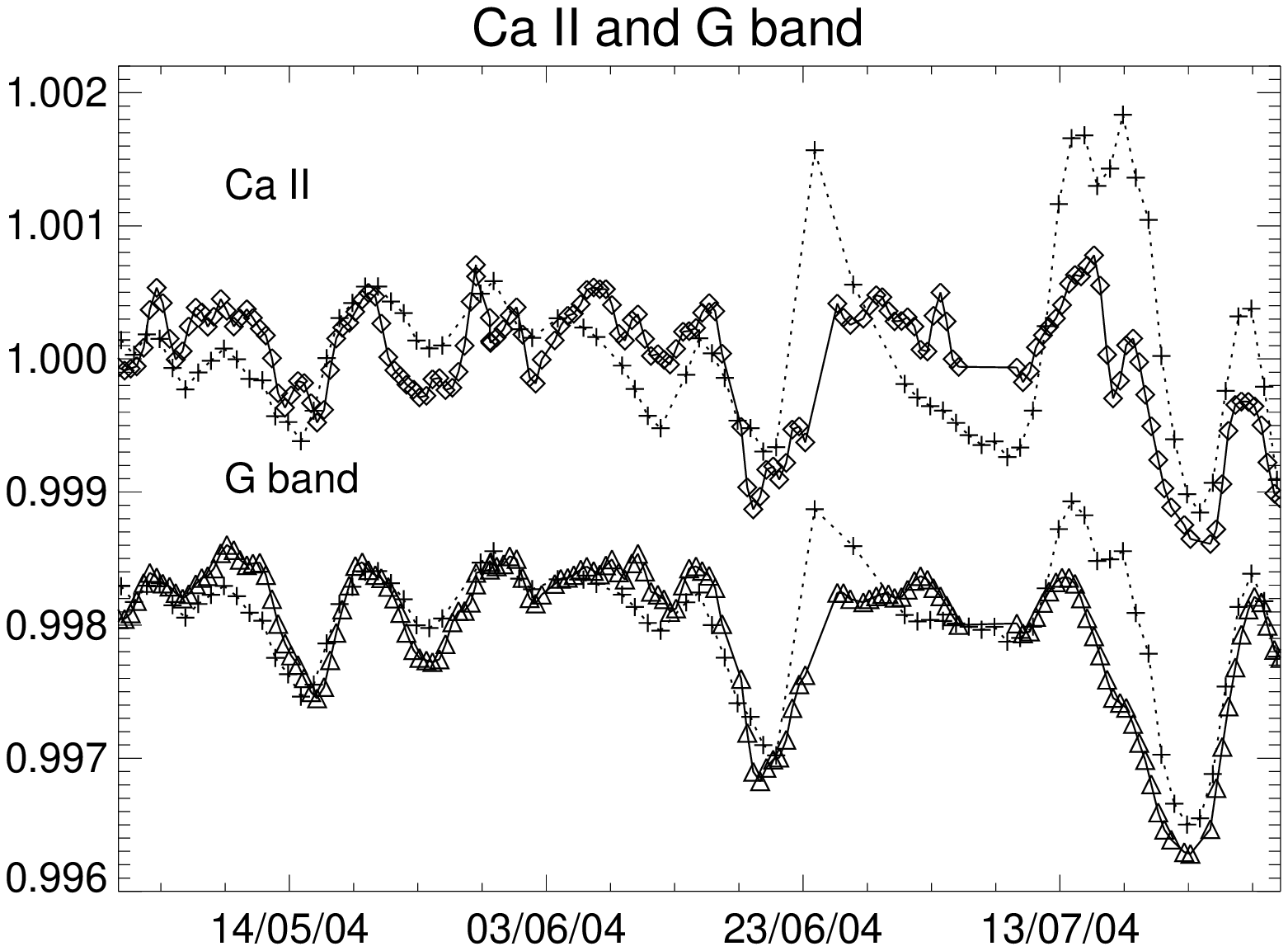}}
        \resizebox{\hsize}{!}{\includegraphics{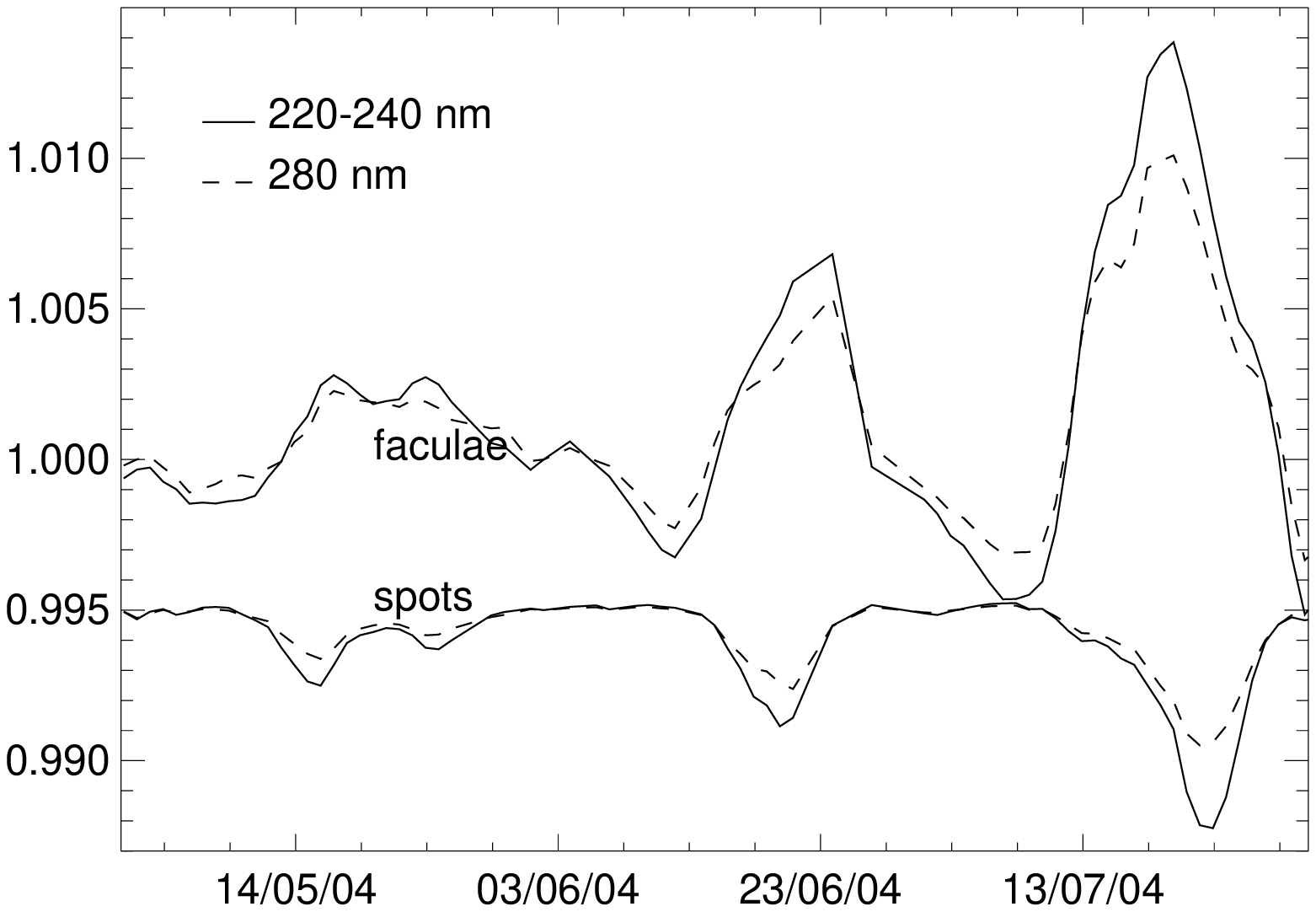}
                              \includegraphics{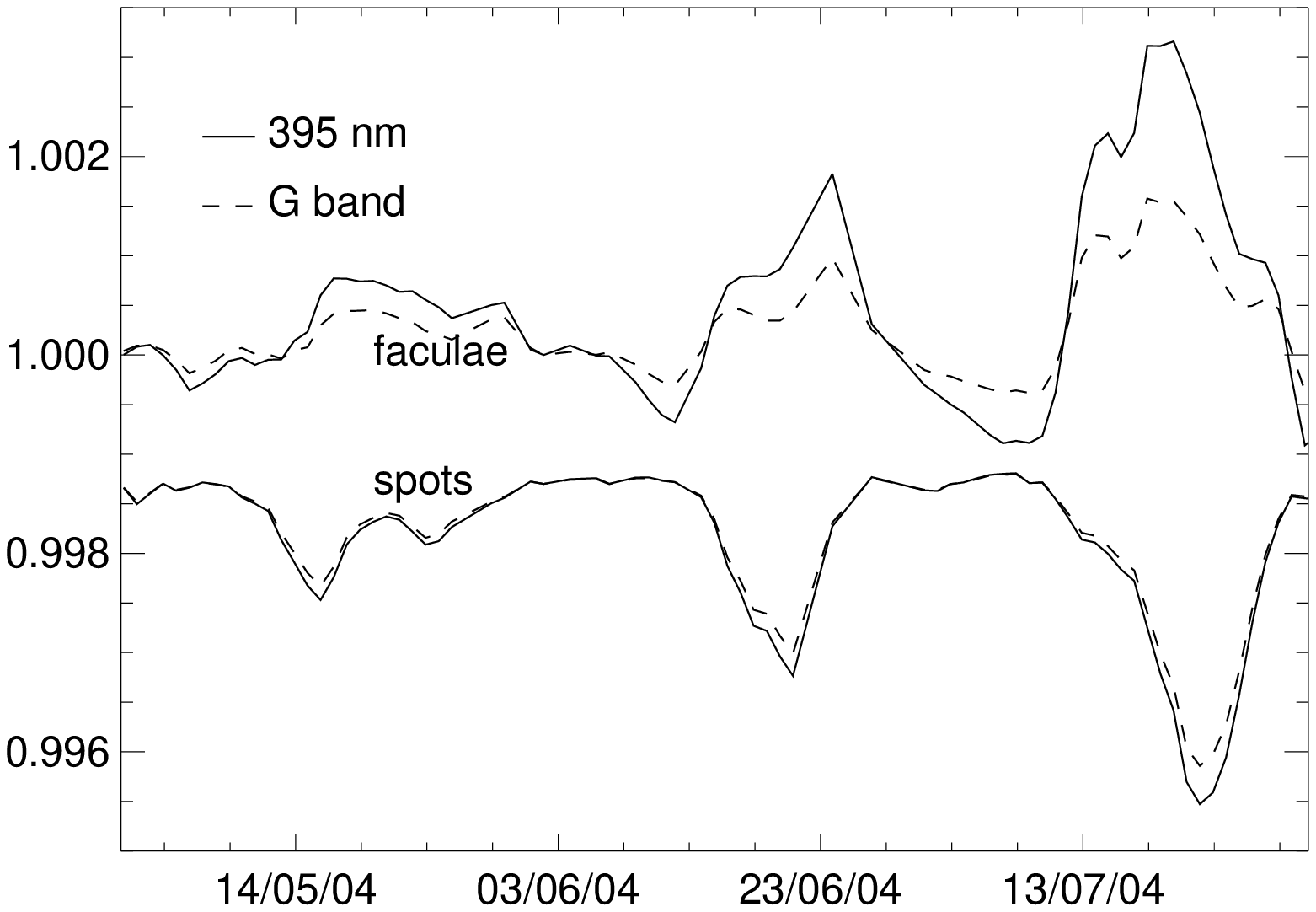}}
\caption[]{Top plots: flux variations for the 230~nm and the Mg~{\sc ii} band
(left-hand side) and of the 395~nm and the G band (right-hand side). The
exact band widths are listed in Tab.~\protect{\ref{tab:bands}}. The diamonds
linked by the solid lines represent SIM data, the plus signs and
dotted lines show the model calculations. Note that the modelled
timeseries have been binned onto daily values; the SIM data remain
on their original time resolution, but have been smoothed using a
binomial filter. Bottom plots: Modelled time series for the facular and spot contributions.
The upper lines are for the faculae, the lower for the spots. The band wavelengths
for the solid and dashed lines are indicated on the plots.
}
\label{fig:timeseries1}
\end{figure*}

%
\subsection{SIM \& SATIRE time series}
\label{sec:timeseries}
In this section, we compare the variability in a number of wavelength 
bands in more detail. The wavelength bands have been picked so as to 
show the change in behaviour going from the UV around 220~nm, up to the IR 
at 1.5~$\mu$m. They are typically also chosen at wavelengths where the 
relative variability is high and the data quality is good. 
The wavelength bands are listed in Tab.~\ref{tab:bands}, along with their
band widths and the number of data points included in the integration
of the SIM data. So as to improve the S/N level of the resulting 
time series, the UV bands, in particular, have been chosen to 
include a large number of wavelength points. Note that the noise level
of the modelled time series is largely governed by the noise in the 
magnetograms and cannot be decreased by increasing the number of 
wavelength points considered; the absolute flux level, however, 
is influenced by the coarseness of the wavelength grid. 

\begin{figure*}
        \resizebox{\hsize}{!}{\includegraphics{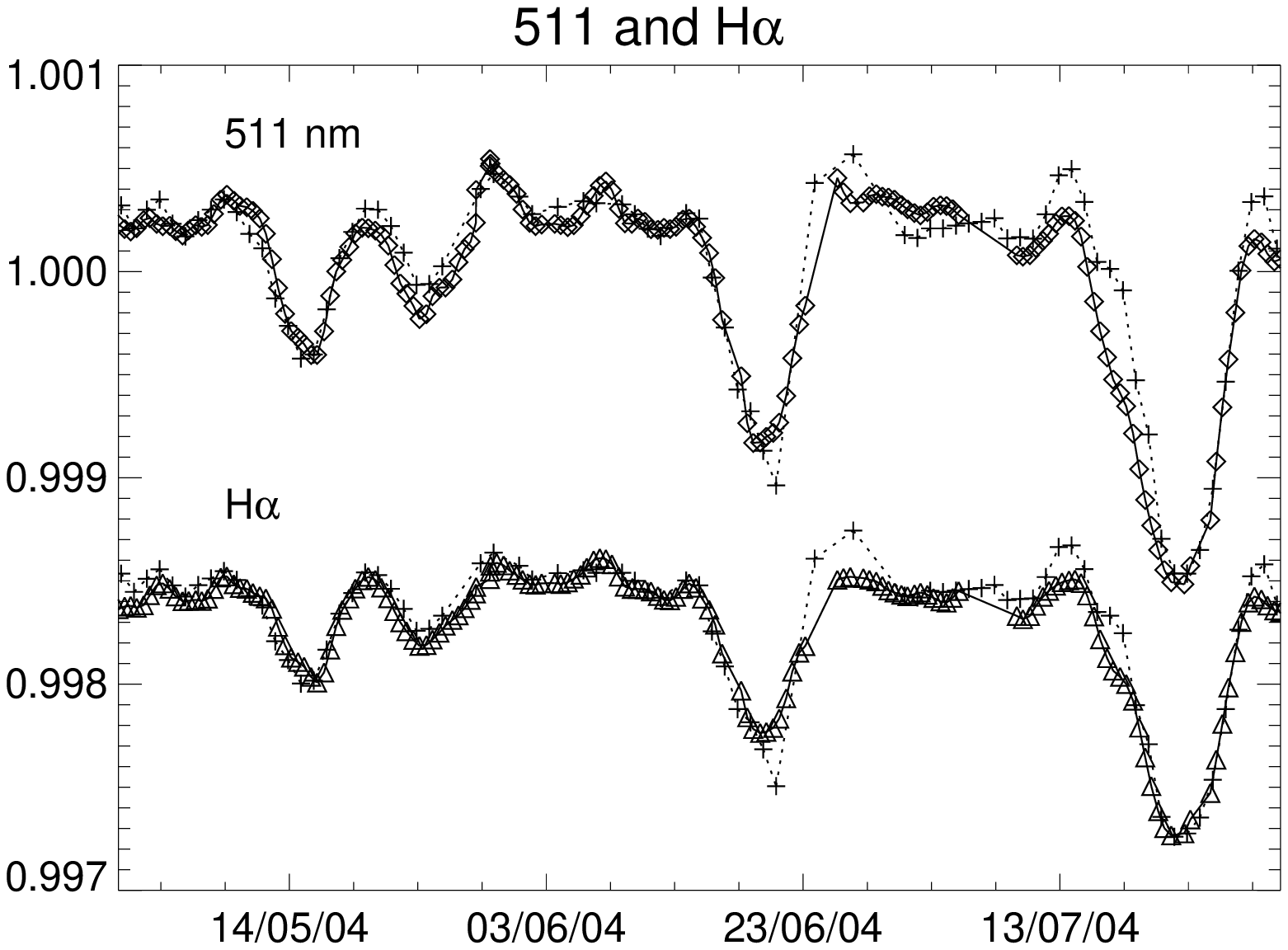}
			      \includegraphics{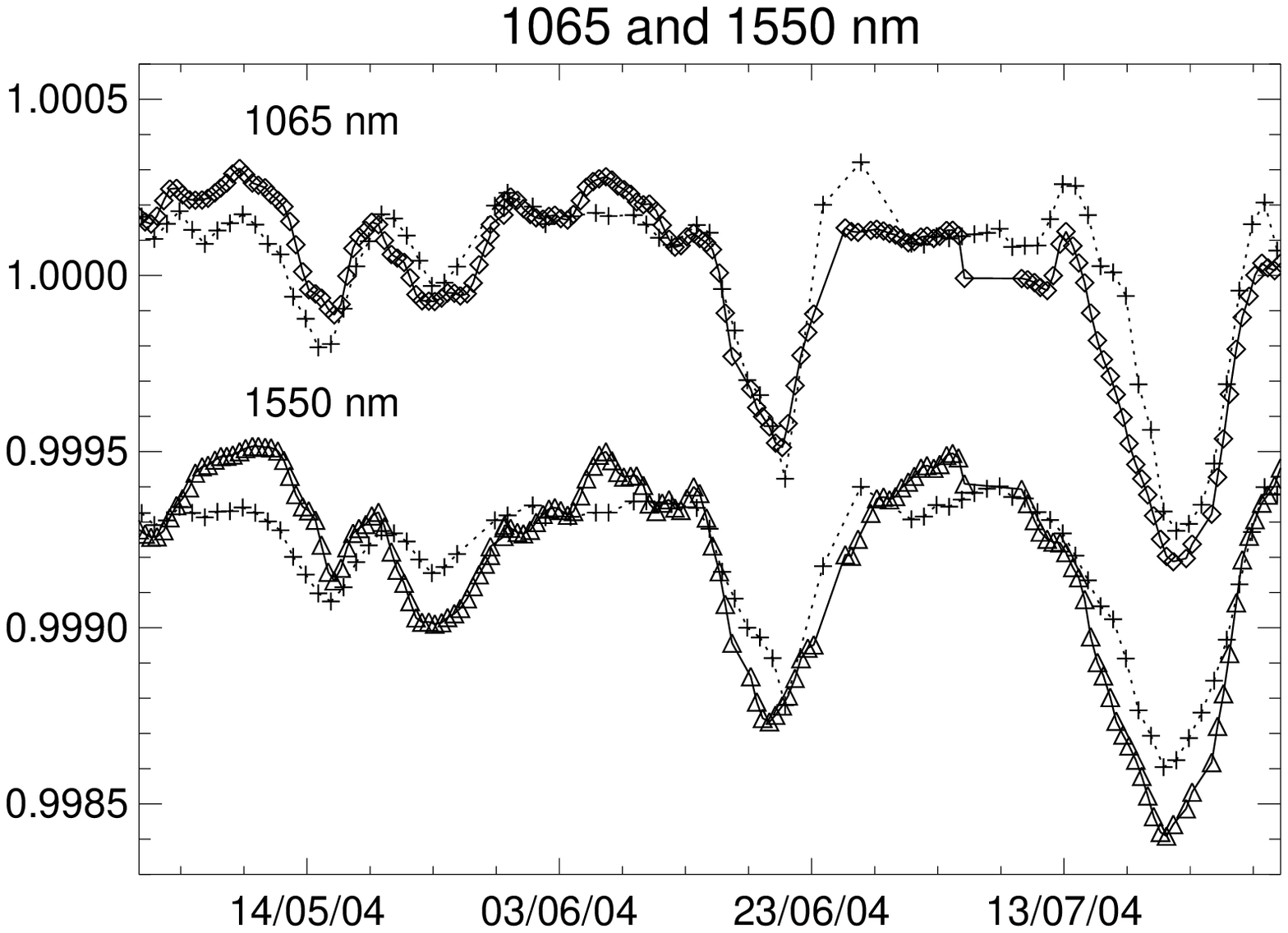}}
        \resizebox{\hsize}{!}{\includegraphics{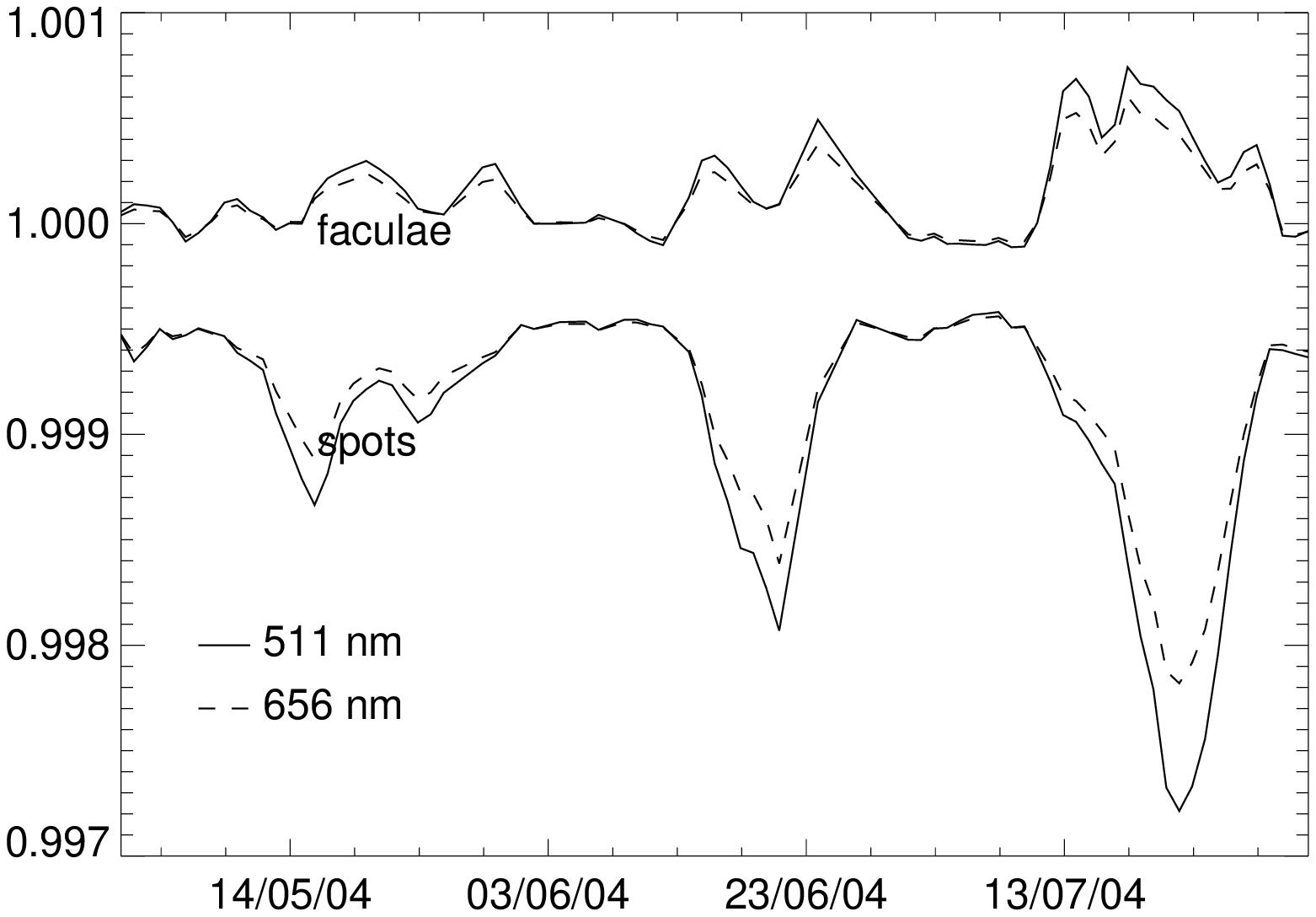}
			      \includegraphics{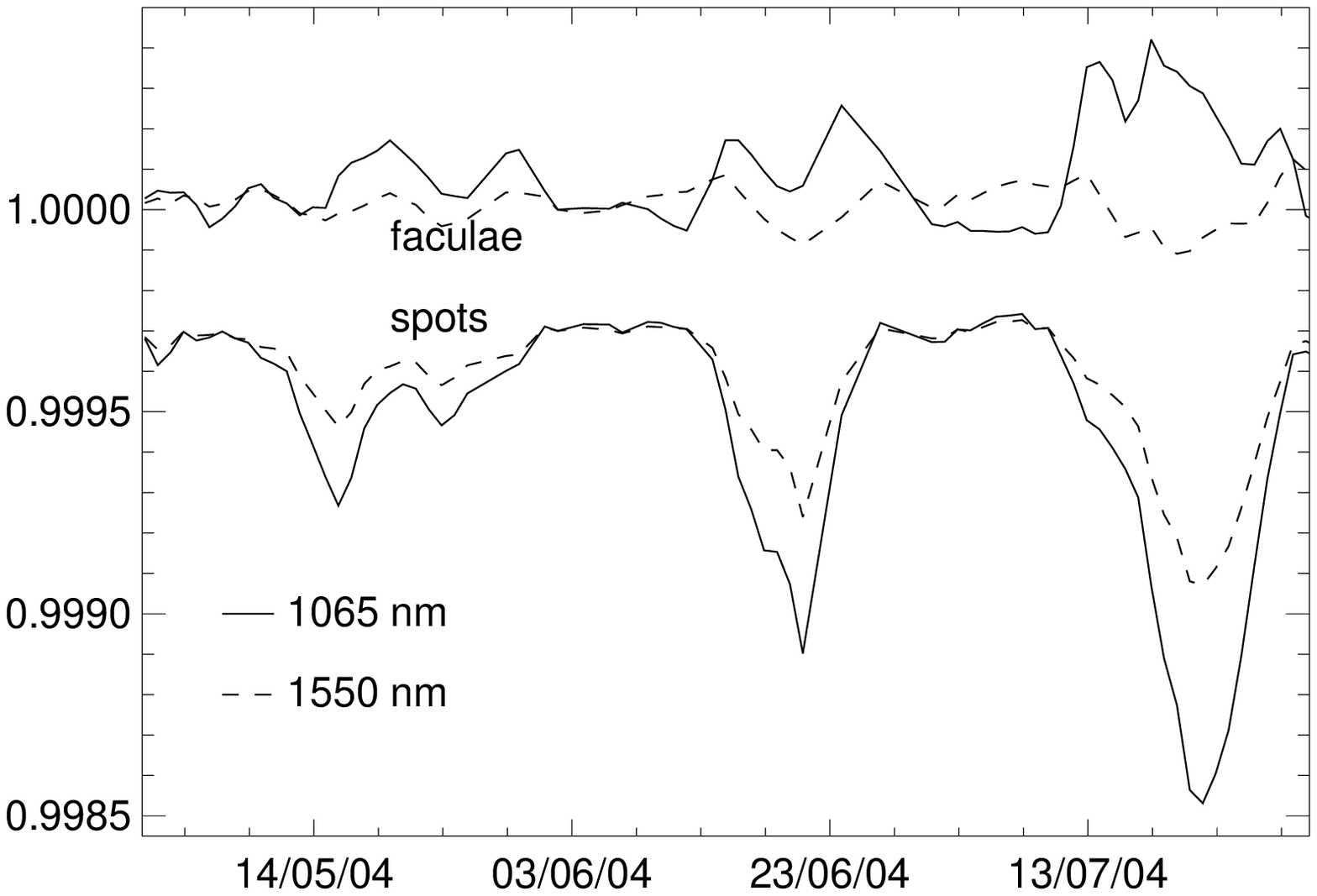}}
\caption[]{Same as Fig.~\protect{\ref{fig:timeseries1}}, though this time 
showing wavelength bands centred at 511~nm and at H$\alpha$ respectively 
(left-hand side), and two near-IR bands centred at 1065 and 1550~nm, 
respectively (right-hand panels). }
\label{fig:timeseries2}
\end{figure*}

The measured and modelled timeseries are shown in Figs~\ref{fig:timeseries1} and 
\ref{fig:timeseries2} and the correlation coefficients as well as the 
slopes between the model and the data are listed in Tab.~\ref{tab:bands}. The 
upper panels show the measured and modelled solar variability while 
the bottom panels show the modelled contributions of the faculae and 
spots to the irradiance variations. The figures illustrate again 
the rapid decrease in the variability towards longer wavelengths. Additionally, 
they show the change in the lightcurve aspects as one moves from the UV 
to the visible. In the UV, as illustrated by the 230 and 280~nm bands 
(Fig.~\ref{fig:timeseries1}), the influence
of the spots is so small that their darkening effect is more than compensated for 
by the faculae, even on solar rotational time scales. Furthermore, the
facular contrast increase towards the limb is not sufficient at these 
wavelengths to counteract the projection effects. Consequently, the Sun appears 
brightest when the main spot groups are nearly at disk centre. 

At our resolution, this behaviour is no longer observed at longer wavelengths, 
such as in the Ca~{\sc ii} and the G bands. The spot darkening is now 
sufficient to offset the facular brightening,
at least when the active regions are near disk centre. The overall lightcurves
thus appear somewhat confusing, with rapid sequences of peaks and dips,
and very few stretches of `quiet'-Sun behaviour. Further high-resolution
calculations would be required to check whether this behaviour also
holds for the individual line cores, or whether we are currently seeing
a mix of line and continuum behaviour. The low wavelength
resolution of the model and the failure to calculate exact line profiles
partly explains the relatively large difference in the observed and modelled mean
flux in the Mg~{\sc ii} and, in particular, 395~nm (Ca H\&K) bands, where the model only
includes 4 wavelength points. A further reason for the difference in flux is due
to our assumption of LTE, most markedly for Ca H\&K.

The situation again changes when looking at
(continuum) bands and longer wavelengths in general (see left-hand panel
in Fig.~\ref{fig:timeseries2}). The facular brightenings
are now much weaker and mostly show a double-peaked aspect, i.e., the faculae
produce most of the brightening when near the limb, and very little or even no
brightening when at disk centre. Combined with the spot contribution, this
leads to the familiar spot-dominated light curves, with small brightenings
just before and after spot passages. Such behaviour is indeed seen for the
TSI, and has been discussed at length in the previous section.

\begin{figure}
        \resizebox{\hsize}{!}{\includegraphics{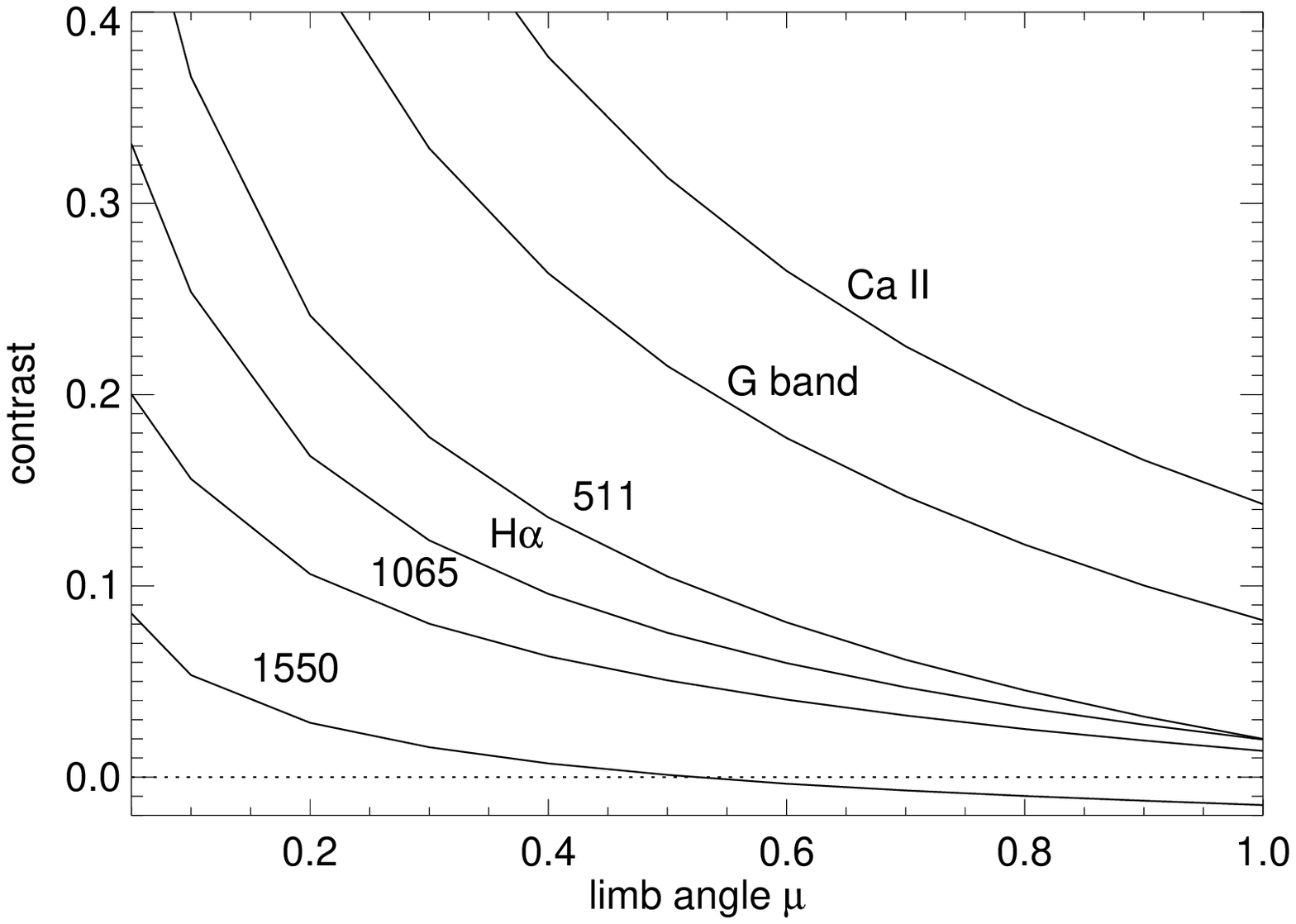}}
\caption[]{A plot of the modelled contrasts in the 6 longer wavelength
bands from Tab.~\protect{\ref{tab:bands}}. The contrast decreases
very strongly with wavelength, becoming negative at the disk centre
at the near IR, as illustrated by the bottom curve for 1550~nm.
The contrasts in the Mg~{\sc ii} and 230~nm bands are much higher and
are not plotted here. Note also that these contrasts are maximum values that are scaled by
the facular filling factors.}
\label{fig:contrasts}
\end{figure}

In the NIR, finally, the facular brightenings become very weak, and might
even disappear completely for the facular model atmosphere we employ. 
This is suggested by the model calculations for the 1550~nm band that 
differ markedly from those for the 1065~nm band. Fig.~\ref{fig:timeseries2} 
suggests that the spots are not sufficiently dark at 1550~nm and indicates 
that the temperature of the spot model atmosphere is too high in the deeper 
layers (1550~nm is close to the opacity minimum and thus carries information 
on the deepest observed layers). But while the model calculations appear 
to underestimate the flux decrease due to most of the active regions, 
they agree very well with the timing of the flux decrease; note that the 
dips due to the spot passages are significantly wider at 1.55 than 
at 1.07~$\mu$m in the observed as well as in the modelled data. In the model, 
the wider spot passages are a consequence of the very small contrast 
of the faculae at that wavelength. Fig.~\ref{fig:contrasts} shows 
a comparison of the model facular contrast for the six longer wavelength 
bands. The contrast at 1550~nm shows the lowest values throughout and 
becomes negative near the disk centre ($\mu > 0.5$). This low contrast
means that only very little facular brightening is seen near the limb and 
thus leads to an earlier onset of the spot-induced darkening, as illustrated 
on the right-hand plots of Fig.~\ref{fig:timeseries2}. 

We note that at 1.55~$\mu$m, there is rather poor agreement between the 
model calculations and the SIM data during May 2004. In particular, we find
that the SIM data show a reversal in the relative spot strength during  
May. The first spot (centred around May 15th) is significantly stronger
than the spot at the end of May in all wavebands, except at 1.55~$\mu$m, where
the second spot appears darker. This could indicate that
the temperature gradient in the two spots is different, although we
cannot exclude uncorrected data fluctuations. 

\begin{table*} 
\caption[]{Table listing the different bands used to calculate 
the time series. The first column gives the label as used for 
the plots, the second and third columns give the start and 
end wavelengths for the bands, while column four gives the number
of flux points over which the integral was carried out for the 
SORCE/SIM data. Columns 5 and 6 give the mean fluxes (in units of
W~m$^{-2}$) of the band as derived from the model calculations and the
SIM measurements. The last three columns list the correlation coefficients,
the gradient and the y-axis flux offsets of a linear fit of the data to the model;
the gradient and offset for Ca~{\sc ii} is 
in brackets as the correlation coefficient is rather low.}
\begin{center}
\begin{tabular}{lrrcccccc}
\hline
band    & $\lambda_S$ \ & $\lambda_E$ \ & number of & flux & flux & correlation & gradient & offset     \\
        & [nm]          & [nm]          & flux points & (data) & (model) & coef & (vs model) &          \\
\ [1]    & [2]           & [3]           & [4]   & [5]           & [6]   & [7]   & [8]      & [9]        \\
\hline 
230             & 220   & 240   & 271   & 1.02  & 0.98  & 0.90  & 0.99 & 0.06 \\
Mg~{\sc ii} h\&k & 277  & 283   & 38    & 1.20  & 0.71  & 0.96  & 1.03 & 0.47 \\
Ca~{\sc ii} H\&K & 391  & 398   & 14    & 7.21  & 5.60  & 0.51  & (0.95) & (1.89) \\
G-band          & 420   & 435   & 22    & 23.5  & 21.3  & 0.77  & 1.35  & -5.3 \\
511             & 507   & 516   &  7    & 13.5  & 15.3  & 0.95  & 0.88  & \ 0.1 \\
H$\alpha$       & 644   & 668   & 10    & 32.0  & 33.8  & 0.96  & 0.84  & \ 3.6 \\
1065            & 1050  & 1080  & 6     & 16.7  & 15.8  & 0.89  & 1.24  & -2.9 \\
1550            & 1527  & 1583  & 10    & 12.9  & 14.6  & 0.93  & 1.33  & -6.4 \\
\hline
\end{tabular}
\end{center}
\label{tab:bands}
\end{table*}

\section{Discussion and Conclusions}
\label{sec:conclusion}

We have presented and compared SATIRE model calculations and measurements 
of spectral solar variability on rotational time scales. The data and
calculations cover a 3-month time span from May to July 2004. In addition,
we also compare modelled and observed time series of the total irradiance
variability and the variability in a number of selected
wavelength bands. Such comparisons are particularly timely as SORCE/SIM is
able to provide unprecedented observations over most of the range starting in
the UV at approximately 220~nm and including the visible as well as the near
infrared up to 1.6~$\mu$m.

We find excellent agreement between the modelled total solar irradiance 
variations and the SORCE/TIM measurements. The absolute value of the 
wavelength-integrated SORCE/SIM measurements is in line with the expected
model fluxes, and its variability agrees well except on a small number of 
days when the data quality was poorer (see Fig.~\ref{fig:TSI}). We find 
correlation coefficients of 0.97 and 0.92 when comparing the modelled 
total solar irradiance with TIM and the wavelength integrated SIM 
measurements, respectively. 

The modelled and measured spectral variability over the three months is summarised 
in Fig.~\ref{fig:var_all} for wavelengths between 220 and 1600~nm. Overall, we find 
good agreement between the model and the observations. Agreement is particularly 
good between 400 and 1300~nm. In the UV, where we also compare the SIM measurements
to UARS/SUSIM, the agreement is somewhat patchy; some 
strong individual lines, such as the Mg~{\sc ii}~h\&k doublet, match very well, 
others, such as Mg~{\sc i} and Ca~{\sc ii}~H\&K, agree only poorly. This is not
too surprising as we use opacity distribution and assume LTE throughout. 
\citet{uitenbroek1995} have shown that NLTE effects can explain much 
of the different behaviour of the Mg~{\sc i} and  Mg~{\sc ii} resonance lines. 
We also note that the resolution of our calculations is insufficient to resolve even the 
strong lines and to capture their complex behaviour. The role of spectral 
resolution in the context of line variability has been discussed, e.g., 
in \citet{white2000issi}.

In the wavelength range between approximately 310~nm and 350~nm, possibly 
even up to 390~nm, the response of both SORCE/SIM and UARS/SUSIM is 
too poor to determine solar variability on the 
rotational time scale. The best estimate of variability at those wavelengths
is currently provided by the SATIRE model. The model calculations allow us to isolate the 
facular and spot contribution. This, together with the light-curves, illustrates
very clearly the change from facular dominated variability at short
wavelengths to spot-dominated variability above approximately 400~nm.

In the visible, the observed and modelled irradiance variability matches well, though 
the decrease in the variability at longer wavelengths appears somewhat steep 
in the model compared to the observations. We find, e.g., that the SATIRE model overestimates
the variability between about 600 and 800~nm by up to 20~\% compared to the SIM 
measurements, while it underestimates the variability around 1.5~$\mu$m by a similar 
amount. Note that for wavelengths between 800 and 1000~nm, the SIM detectors suffer 
from temperature-induced variability that cannot yet be fully compensated for; we 
were therefore unable to carry out meaningful comparisons at those wavelengths.
As a cross-check, we further compared the SATIRE and SORCE/SIM variability with 
the VIRGO/SPM measurements at 400, 500 and 860~nm. We found that the correlation 
coefficients for the model-to-data comparisons are typically very similar to 
those obtained for the SPM-to-SIM data comparisons. 

Our model suggests that the overall effect of faculae at 1.6~$\mu$m is one of 
darkening, though they appear bright at all wavelengths when seen close to 
the limb. Near 1.6~$\mu$m, the small brightness enhancement seen for faculae
at the limb, however, is typically offset by the spots, so that active-region 
passages produce longer-lasting brightness dips at this wavelength than at 
shorter wavelengths. This is illustrated on the right-hand panel of 
Fig.~\ref{fig:timeseries2}. Contrary to \citet{fontenla2004sorce}, we find 
no evidence for overall bright faculae during the comparatively 
quiet period analysed here. An unambiguous contrast determination is 
difficult, however, as most large facular regions tend to be accompanied
by dark spots whose exact contrasts are also unknown. 

\section*{ACKNOWLEDGMENTS}
The authors would like to thank L.~Floyd for helpful discussions and 
information on SUSIM data, M.~Snow for information on the SORCE/SOLSTICE 
data and C.~Fr\"ohlich for his comments on the paper and help with the 
VIRGO/SPM data. This work was supported by the
\emph{Deut\-sche For\-schungs\-ge\-mein\-schaft, DFG\/} project
number SO~711/1-1 and by the NERC SolCli consortium grant.

\bibliographystyle{aa}


\end{document}